\documentclass[11pt,preprintnumbers,aps,prd,amssymb,nofootinbib,amsmath,superscriptaddress,notitlepage]
{revtex4-2}
\usepackage{epsfig,epsf}
\usepackage{bm} % puts greek and math symbols in boldface using \bm
\usepackage[normalem]{ulem}
\usepackage{color} % {\color{red} ... }
\usepackage{slashed}
\usepackage{relsize}	%rescalable math 
\usepackage{soul} %spac­ing out (let­terspac­ing), un­der­lin­ing, strik­ing out, etc.
\usepackage{hyperref}
\usepackage{tensor} %\indices{^a_b} produces the properly spaced the tensor indices 
\graphicspath{{./figures}}
\newcommand{\beq}{\begin{equation}}
\newcommand{\beql}[1]{\begin{equation}\label{#1}}
\newcommand{\eeq}{\end{equation}}
\def\bal#1\gal{\begin{align}#1\end{align}}

\newcommand{\eq}[1]{(\ref{#1})}

\renewcommand{\b}[1]{{\bm #1}} 
\newcommand{\unit}[1]{\hat {{\bm #1}}} % unit vector

\newcommand{\mean}[1]{\langle #1 \rangle}

\newcommand{\D}{{\rm d}}

\newcommand{\I}{{\rm i}}

\newcommand{\h}[1]{\widehat{#1}}

\newcommand{\E}{{\rm e}}
\newcommand{\p}{{\rm p}}

\newcommand{\de}{\partial}
\renewcommand{\vec}[1]{\ensuremath{\mathchoice				%   vec: bold vector
		{\mbox{\boldmath$\displaystyle\mathbf{#1}$}}
		{\mbox{\boldmath$\textstyle\mathbf{#1}$}}
		{\mbox{\boldmath$\scriptstyle\mathbf{#1}$}}
		{\mbox{\boldmath$\scriptscriptstyle\mathbf{#1}$}}}}

\begin{document}

\title{Causal fermion states in magnetic field in a relativistic rotating frame and electromagnetic radiation by a rapidly rotating charge}

\author{Matteo Buzzegoli}

\author{Kirill Tuchin}

\affiliation{
Department of Physics and Astronomy, Iowa State University, Ames, Iowa, 50011, USA}

%\date{\today}

\begin{abstract}

We consider the Dirac field uniformly rotating with angular velocity $ \Omega$ and also subject to the constant magnetic field $\b B$ directed along the rotation axis. 
The causal states are constrained to the interior of the light cylinder of radius $c/\Omega$. When this radius is smaller than the system size, as in the quark-gluon plasma, the effect of the boundary on the fermion spectrum is critical. We derive the fermion spectrum and study its properties. We compute the intensity of the electromagnetic radiation emitted due to transitions between the fermion states. We study its dependence on energy and angular momentum for different values of the angular velocity and the magnetic field. Rotation has enormous impact on the electromagnetic radiation by the quark-gluon plasma with or without the magnetic field. 

\end{abstract}

\maketitle

%%%%%%%%%%%%%%%%%%%%%%%%%%%%%%%%%%%%%%%%
\section{Introduction}\label{sec:a}

Our recent articles \cite{Buzzegoli:2022dhw,Buzzegoli:2023vne} initiated study of the electromagnetic radiation emitted by an electrically charged particle embedded into the rotating medium placed in the external magnetic field $B$. We observed that it exhibits a remarkable sensitivity to the angular velocity of rotation $\Omega$: even relatively slow rotation changes the radiation spectrum at high energies of the emitting particle. Thus it can be used as a powerful tool to study any rapidly rotating systems. However, our main motivation is to understand the electromagnetic spectrum emitted by the quark-gluon plasma.  Previous studies of the rotating quantum systems made significant progress in understanding their  thermodynamic and kinetic properties \cite{Vilenkin:1980zv,Becattini:2007nd,Ambrus:2014uqa,Ambru2021,McInnes:2014haa,McInnes:2015kec,Ambrus:2015lfr,Chen:2015hfc,McInnes:2016dwk,Jiang:2016wvv,Chernodub:2016kxh,Ebihara:2016fwa,Chernodub:2017ref,Liu:2017spl,Buzzegoli:2017cqy,Zhang:2018ome,Buzzegoli:2020ycf,Chen:2019tcp,Wang:2018zrn,Wang:2019nhd,Palermo:2021hlf,Sadooghi:2021upd}. Quantization of rotating quantum field was discussed before in \cite{Letaw:1979wy,Bakke:2013sla,Konno:2012rt,Ayala:2021osy,Manning:2015sky} and a variety of peculiar effects associated with rotation were considered in \cite{Anandan:1992zz,POST:1967qwl,Aharonov1973-AHAQAO,Anandan:1977ra,Staudenmann:1980uqe,PhysRevLett.106.076601,Fonseca:2017pnk,Chen:2021aiq,Chernodub:2017ref}. The effect of rotation and magnetic field on bound states was also recently addressed in \cite{Tuchin:2021lxl,Tuchin:2021yhy,Buzzegoli:2022omv}.

In \cite{Buzzegoli:2022dhw,Buzzegoli:2023vne} we assumed that the magnetic field is constant and the rotation of the fermions is uniform. This allowed us to employ the well-known solutions of the Dirac equation in the magnetic field in cylindrical coordinates to obtain the fermion wave functions in the rotating frame. We then computed the intensity of the electromagnetic radiation in the laboratory frame using the leading order of the perturbation theory. We explicitly made another important assumption. 
In the reference frame uniformly rotating along with the fermions with angular velocity $\Omega$, the radial extent of the spacetime is bound by the condition $r\le R=1/\Omega$, where $r$ is the distance from the rotation axis. The radially moving light signal cannot cross the $r=R$ boundary. However, the equations of motion in the rotating frame do not explicitly restrict the solutions to the allowed region. This necessitates imposing the boundary conditions at the spacetime radial boundary $R$. In \cite{Buzzegoli:2022dhw,Buzzegoli:2023vne} we avoided this complication, by considering what we termed ``the (relatively) slow rotation,'' which refers to the angular velocities and magnetic field strengths such that $\sqrt{qB}\gg \Omega$. Since the extent of the fermion wave function in the radial direction is of the order of the magnetic length $1/\sqrt{qB}$, it is exponentially suppressed at the spacetime boundary in this approximation. This allowed us to ignore the boundary conditions and let the solutions formally extend to infinity in the radial direction. Whereas this approximation is exact in nearly all physical systems, it fails in the quark-gluon plasma and possibly in the extremal black holes.

In the present paper we extend the results of \cite{Buzzegoli:2022dhw,Buzzegoli:2023vne}  beyond the slow rotation approximation by imposing the appropriate boundary conditions on the solutions to the Dirac equation at $r=1/\Omega$. The importance of the causal boundary was emphasized in a number of papers   that studied the statistical properties of rotating quantum systems \cite{Duffy:2002ss,Ebihara:2016fwa,Chernodub:2017ref,Chernodub:2017mvp}. In \cite{Buzzegoli:2022omv} we studied the effect of rotation, including the causal boundary, on the bound states. 
Our main goal in this paper is to derive the intensity of the electromagnetic radiation taking the boundary condition into account. 

The paper is structured as follows. In Sec.~\ref{sec:b} we discuss the exact solution of the Dirac equation for rotating fermions in a constant magnetic field in a finite cylinder of radius $R=1/\Omega$ without specifying the boundary condition (BC). In Sec.~\ref{sec:unboundBC} we consider the limiting case $R\to \infty$ and show how the well-known formulas of Sokolov and Ternov \cite{Sokolov:1986nk} are recovered. In Sec.~\ref{sec:MITBC} we apply the MIT boundary condition to the general solutions of the Dirac equation and discuss the resulting energy spectrum. The limit of vanishing magnetic field is studied in 
Sec.~\ref{sec:JustRotation}. The differential and total radiation intensity of the radiation emitted by these causal states is derived in Sec.~\ref{sec:RadiationInt}. Our main analytical result is the differential radiation intensity in Eq. (\ref{eq:DiffIntensityCylinder}). The numerical procedure is described in Sec.~\ref{sec:Results}. The total intensity as a function of energy for different angular velocities $\Omega$ is displayed in Fig.~\ref{fig:EnergydepMITBC}. The total intensity as a function of angular velocity $\Omega$ at fixed energy is shown in Fig.~\ref{fig:WVsRot} and the angular momentum distribution of the radiation is exhibited in Fig.~\ref{fig:PAMVsRot}. We summarize in Sec.~\ref{sec:summary}.  The natural units $\hbar=c=1$ are used throughout. 

%%%%%%%%%%%%%%%%%%%%%%%%%%%%%%%%%%%%%%%%
\section{General solution to Dirac equation in magnetic field in rotating frame}\label{sec:b}

It is instructive to begin with a brief review  of the Dirac equation  in the frame rotating with the constant angular velocity $\b\Omega=\Omega\unit z$. The constant magnetic field $\b B= B\unit z$ is conveniently described in the symmetric gauge $A^\mu =(0, -B y/2, B x/2,0)$. The Dirac Hamiltonian $H$ in the rotating frame is obtained from that in the stationary frame by adding the operator $-\Omega J_z$, where $J_z= -i\partial_\phi+\frac{i}{2}\gamma^x\gamma^y$ is the angular momentum projection on the rotating axis:
\begin{align}
\label{eq:H}
H= \gamma^0\b \gamma\cdot (\b p-q\b A)+\gamma^0M-\Omega J_z\,.
\end{align}
In view of the axial symmetry, we employ cylindrical coordinates $r$, $\phi$, $z$. In place of the radial coordinate $r$ it will be advantageous to use a dimensionless variable $\rho= \frac{|qB|}{2}r^2$. The light-cylinder is located at $\rho=\rho_R$, where
\begin{equation}\label{eq:RhoR}
 \rho_R=\frac{|qB|R^2}{2}=\frac{|qB|}{2\Omega^2}.   
\end{equation}
We are seeking a solution to the Dirac equation in the standard representation in the form 
\begin{equation}
\label{eq:SolutionDiracForm}
\psi(t,r,\phi,z)=\E^{-\I\epsilon E t}\frac{\E^{\I p_z z}}{\sqrt{2\pi}}\frac{\E^{\I m \phi}}{\sqrt{2\pi}}  \left(\begin{array}{c}
	f_1(\rho) \E^{-\I\phi/2}\\
	f_2(\rho)\E^{+\I\phi/2} \\
	f_3(\rho)\E^{-\I\phi/2} \\
	f_4(\rho)\E^{\I\phi/2}
\end{array}\right)\,,
\end{equation}
where $p_z$ is the momentum component along the rotation axis, $\epsilon=+/-$ labels 
particle/antiparticle wave functions, integer $m$ is the magnetic quantum number
and the radial functions $f_s(\rho)$, $s=1,\ldots,4$ must satisfy the boundary condition that we specify in the next section.
It is convenient to define the operators \cite{Sokolov:1986nk}:
\begin{align}\label{eq:DefR1R2}
R_1 =& \sqrt{\frac{|qB|}{2}\rho}\left(2\frac{\D}{\D\rho}+\bar\sigma-\frac{m-\tfrac{1}{2}}{\rho} \right),\quad 
R_2 = \sqrt{\frac{|qB|}{2}\rho}\left(2\frac{\D}{\D\rho}-\bar\sigma+\frac{m+\tfrac{1}{2}}{\rho} \right),
\end{align}
where $\bar{\sigma}=\text{sgn}\left(qB\right)$.
Substituting (\ref{eq:SolutionDiracForm}) into the Dirac equation and using \eq{eq:DefR1R2} one finds the equations for the radial functions:
\begin{equation}
\label{eq:2ndOrderRadialDiracEq}
R_1 R_2 f_{2,4}(\rho) = -2 |qB|\lambda f_{2,4}(\rho),\quad
R_2 R_1 f_{1,3}(\rho) = -2 |qB|\lambda f_{1,3}(\rho),
\end{equation}
and the dispersion relation 
\begin{equation}\label{dispersion}
    2|qB|\lambda = (\epsilon E - \Omega\, m)^2 - M^2 - p_z^2\,,
\end{equation}
where $M$ is the fermion mass and the real parameter $\lambda$ is the principal quantum number.

The solutions of (\ref{eq:2ndOrderRadialDiracEq}), finite at the origin, are
\begin{align}\label{eq:solutionsRadial}
f_{1,3}^{\bar\sigma} = & \sqrt{|qB|}C^{\bar\sigma}_{1,3} F^{\bar\sigma}_1(\rho),\quad 
f_{2,4}^{\bar\sigma} = \sqrt{|qB|}\I C^{\bar\sigma}_{2,4} F^{\bar\sigma}_2(\rho),
\end{align}
where $F_{1,2}$ are given in terms of the confluent hypergeometric function:
\begin{subequations}
\label{eq:DefHyperF}
\begin{align}
F^{\bar\sigma}_1(\rho) =& \frac{N_1}{\Gamma(m+\frac{1}{2})}\, \E^{-\rho/2} \rho^{\frac{m-\tfrac{1}{2}}{2}}
    \,_1F_1\left(-\left(\lambda-\frac{1-\bar\sigma}{2}m-\frac{1-\bar\sigma}{2}\frac{1}{2}\right);\, m+\frac{1}{2};\, \rho \right),\\
F^{\bar\sigma}_2(\rho) =&  \frac{N_2}{\Gamma(m+\frac{3}{2})}\, \E^{-\rho/2} \rho^{\frac{m+\tfrac{1}{2}}{2}}
    \,_1F_1\left(-\left(\lambda-\frac{1-\bar\sigma}{2}m-\frac{3+\bar\sigma}{4}\right);\, m+\frac{3}{2};\, \rho \right),
\end{align}
\end{subequations}
and  $N_1$ and $N_2$ are the normalization constants such that
\begin{equation}
\begin{split}
\int_0^{\rho_R} \D\rho\,\left[F^{\bar\sigma}_1(\rho)\right]^2 =& 1,\quad
\int_0^{\rho_R} \D\rho\,\left[F^{\bar\sigma}_2(\rho)\right]^2 = 1.
\end{split}
\end{equation}
 In general, $N_1$ and $N_2$ depend on $m,\,\lambda,\,\bar\sigma$ and $R$.
The gamma factors in (\ref{eq:DefHyperF}) ensures that the functions are regular and real for every value of $m$ and $\rho$ \cite{NISTHandbook}.
Using (\ref{eq:DefR1R2}) and (\ref{eq:DefHyperF}) we can also show that
\begin{equation}
\label{eq:R1R2action}
R_1 F^{\bar\sigma}_1(\rho) = - \sqrt{2|qB|} \lambda \frac{N_1}{N_2} F^{\bar\sigma}_2(\rho), \quad
R_2 F^{\bar\sigma}_2(\rho) = + \sqrt{2|qB|} \frac{N_2}{N_1} F^{\bar\sigma}_1(\rho).
\end{equation}
Employing these formulas it is straightforward to demonstrate that $F_{1,2}$ solve
the Eqs. (\ref{eq:2ndOrderRadialDiracEq}).
In summary, the general form of solutions to the Dirac equation 
in the magnetic field in rotating frame is
\begin{equation}
\label{eq:SolutionDiracGeneral}
\psi^{\rho_R}(t,r,\phi,z)=\E^{-\I\epsilon E t}\frac{\E^{\I p_z z}}{\sqrt{2\pi}}\frac{\E^{\I m \phi}}{\sqrt{2\pi}} \sqrt{|qB|} \left(\begin{array}{c}
	C^{\bar\sigma}_1 F^{\bar\sigma}_1(\rho) \E^{-\I\phi/2}\\\I C^{\bar\sigma}_2 F^{\bar\sigma}_2(\rho)\E^{+\I\phi/2} \\
	C^{\bar\sigma}_3 F^{\bar\sigma}_1(\rho)\E^{-\I\phi/2} \\ \I C^{\bar\sigma}_4 F^{\bar\sigma}_2(\rho)\E^{\I\phi/2}
\end{array}\right),
\end{equation}
with $F_{1,2}^{\bar \sigma}$ given by \eq{eq:DefHyperF}. $C^{\bar \sigma}_s$, $s=1,\ldots,4$ are constant coefficients. 

It is remarkable that eigenfunctions \eq{eq:SolutionDiracGeneral} and their derivatives are continuous across the light cylinder $r=1/\Omega$. This is a general feature of motion in the rotating frame. For example, a free classical particle in the rotating frame spirals out to infinity oblivious of the light cylinder. Of course, the very definition of the rigidly rotating frame breaks down at the light cylinder. In Sec.~\ref{sec:MITBC} we cut off the physically meaningless region by imposing an appropriate boundary condition.

%%%%%%%%%%%%%%%%%%%%
%%%%%%%%%%%%%%%%%%%%
\subsection{Transverse polarization states}\label{sec:trans.pol.}
%%%%%%%%%%%%%%%%%%%
%%%%%%%%%%%%%%%%%%%%%%
The constants $C_s$ are not completely fixed by the Dirac equation, which only requires
that the constants satisfy the algebraic system
\begin{subequations}
\label{eq:DiracEqCBounded}
\begin{align}
(\varepsilon E -\Omega\, m \mp M) C^{\bar\sigma}_{1,3} - \frac{N_2}{N_1} \sqrt{2|qB|}C^{\bar\sigma}_{4,2} - p_z C^{\bar\sigma}_{3,1} =& 0 , \\
(\varepsilon E -\Omega\, m \mp M) C^{\bar\sigma}_{2,4} - \lambda\frac{N_1}{N_2} \sqrt{2|qB|}C^{\bar\sigma}_{3,1} + p_z C^{\bar\sigma}_{4,2} =& 0 .
\end{align}
\end{subequations}
Furthermore, imposing the normalization of the wave function function as
\begin{equation}
\int \psi^\dagger \psi \D^3 x =1,
\end{equation}
we find the constraint
\begin{equation}
\label{eq:NormBounded}
\sum_{s=1}^4 C^{\bar\sigma\, *}_s C^{\bar\sigma}_s = 1.
\end{equation}
To fix the constants we can require the fermions to be in a certain polarization state so that the wave functions \eq{eq:SolutionDiracGeneral} are
the corresponding  eigenfunctions. It is also convenient to choose a polarization
operator which commutes with the Hamiltonian. As in the case of unbounded solutions, i.e.\ when $R\to \infty$, a
commuting polarization operator is given by the spin magnetic moment%\cite{book:SokolovAndTernov,book:Bordovitsyn}
\begin{align}\label{eq:def-mu}
\vec{\mu}= \vec{\Sigma}
    - \frac{\I\gamma_0 \gamma_5}{2}\vec{\Sigma}\times (\vec{p}-q\vec{A})\,.
\end{align}
The spin magnetic moment component along the direction of the magnetic spin is a constant of motion.
Eigenstates of this operator are said to have the transverse polarization
and are obtained by solving the eigenvalue problem
\begin{equation}
\h{\mu}_z \,\psi = \zeta\, \sqrt{(E - \Omega \, m)^2 -p_z^2}\; \psi,
\end{equation}
with $\zeta=\pm 1$. This is equivalent to the set of equations for the constants $C_s$
\begin{equation}
\label{eq:transverse1}
\begin{split}
\left[(\varepsilon E - \Omega\, m)\mp \zeta \bar{E}\right] C^{\bar\sigma}_{1,3} = & p_z C^{\bar\sigma}_{3,1}, \\
\left[(\varepsilon E - \Omega\, m)\pm \zeta \bar{E}\right] C^{\bar\sigma}_{2,4} = & -p_z C^{\bar\sigma}_{2,4},
\end{split}
\end{equation}
where $\bar{E}=\sqrt{(\varepsilon E - \Omega\, m)^2-p_z^2}$. The solution of (\ref{eq:transverse1}) can be written in the form
\begin{equation}
\label{eq:transverse2}
\left(\begin{array}{c}C^{\bar\sigma}_1 \\ C^{\bar\sigma}_2 \\ C^{\bar\sigma}_3 \\ C^{\bar\sigma}_4\end{array}\right)
	= \frac{1}{2\sqrt{2}}
	\left(\begin{array}{c} B_3(A_3 + A_4) \\ B_4(A_4 - A_3) \\ B_3(A_3 - A_4) \\ B_4(A_4 + A_3) \end{array}\right)
\end{equation}
such that
\begin{equation}
\label{eq:transverse3}
\frac{A_3+A_4}{A_3-A_4} = \frac{p_z}{(\varepsilon E - \Omega\,m)-\zeta \bar{E}}.
\end{equation}
The coefficients are now obtained by solving Eqs.  (\ref{eq:DiracEqCBounded}) and imposing the normalization condition (\ref{eq:NormBounded}).
From (\ref{eq:DiracEqCBounded}), and reminding that $\lambda$ is non-negative, see Eq. (\ref{dispersion}), we obtain
\begin{equation}
\begin{split}
\frac{B_4}{B_3}
=& \frac{\sqrt{\lambda}N_1}{N_2}\zeta \sqrt{\frac{\bar{E}-\zeta M}{\bar{E}+\zeta M}}
= \frac{\sqrt{\lambda}N_1}{N_2}\zeta\frac{\sqrt{2|qB|\lambda}}{|\bar{E}+\zeta M|}.
\end{split}
\end{equation}
Imposing (\ref{eq:NormBounded}) the coefficients read:
\begin{equation}
\label{eq:CoefficientsTransversePol1Bounded}
C^{\bar\sigma}_{1,3}=\frac{1}{2\sqrt{2}}B_+(A_+ \pm \zeta A_-),\quad
C^{\bar\sigma}_{2,4}=\frac{1}{2\sqrt{2}} \frac{\sqrt{\lambda}N_1}{N_2}
    B_- (A_- \mp \zeta A_+),
\end{equation}
where 
\begin{equation}
\begin{split}
\label{eq:CoefficientsTransversePol2Bounded}
A_\pm = & \left(\frac{E-\Omega\, m \pm p_z}{E-\Omega\, m} \right)^{\frac{1}{2}},\\
B_+ = & \left(\frac{2(\bar{E}+\zeta M)}{(1+\frac{\lambda N_1^2}{N_2^2})\bar{E} +(1-\frac{\lambda N_1^2}{N_2^2})\zeta M} \right)^{\frac{1}{2}},\\
B_- = & \left(\frac{2(\bar{E}-\zeta M)}{(1+\frac{\lambda N_1^2}{N_2^2})\bar{E} +(1-\frac{\lambda N_1^2}{N_2^2})\zeta M} \right)^{\frac{1}{2}}.
\end{split}
\end{equation}
The upper signs in (\ref{eq:CoefficientsTransversePol1Bounded}) refer to $s=1,\,2$ and the lower ones to $s=3,\,4$.

%*********************************************************************************************************
\section{Solutions to Dirac equation with boundary condition at \texorpdfstring{$R\to \infty$}{infinite radius}}\label{sec:unboundBC}
%*********************************************************************************************************
As we mentioned in the Introduction, when the angular velocity of rotation is so small that $\Omega\ll  \sqrt{|qB|}$ one can ignore the exponentially small value of the wave function at $r=R$ and consider the light cylinder as located infinitely far from the rotation axis $r\to \infty$. We refer to this boundary condition and its results as the unbounded case. The boundary condition then requires the vanishing of all components of $\psi$ at $\rho\to \infty$. This can only be realized if the radial quantum number 
\begin{align}\label{def.a}
a=n+\bar\sigma\, m-\frac{1}{2}\,
\end{align}
is a non-negative integer \cite{Sokolov:1986nk,NISTHandbook}. We denoted the principal quantum number $\lambda=n=0,1,\ldots $ to emphasize that it  must be a non-negative integer.
At these values of the radial quantum quantum number, the hypergeometric functions \eq{eq:DefHyperF} reduce, up to a phase factor, to the Laguerre functions
\begin{align}\label{Lagguerre.funct}
I_{n,a}(\rho)=\sqrt{\frac{a!}{n!}} \E^{-\rho/2} \rho^{\tfrac{n-a}{2}} L_a^{n-a}(\rho).
\end{align}
In particular, by multiplying the wave function (\ref{eq:SolutionDiracGeneral}) by an appropriate phase factor, we find from the definitions \eq{eq:DefHyperF}
\begin{equation}
\label{eq:FInf}
\lim_{R\to\infty} F_1 = -\bar\sigma I^{\bar\sigma}_1(\rho),\quad
\lim_{R\to\infty} F_2 = I^{\bar\sigma}_2(\rho),
\end{equation}
where 
\begin{align}\label{Lagguerre}
I^{\bar\sigma}_1(\rho) = I_{n-\frac{1-\bar\sigma}{2},a}(\rho),\quad
I^{\bar\sigma}_2(\rho) =  I_{n-\frac{1+\bar\sigma}{2},a}(\rho).
\end{align}
We now show that this solution to the Dirac equation in constant magnetic field in stationary frame is the same as the one obtained in \cite{Sokolov:1986nk}. For that purpose we move the phase $-\bar{\sigma}$ from $F_1$ in (\ref{eq:FInf}) to a newly defined coefficients $\tilde{C}_{1,3}^{\bar\sigma}=-\bar{\sigma}C_{1,3}^{\bar\sigma}$, such that the eigenfunctions of the Dirac equation read 
\begin{equation}
\label{eq:SolutionDirac}
\psi^\infty_{n,\,a,\,p_z,\,\zeta}(t,r,\theta,\phi)=\E^{-\I\epsilon E t}\frac{\E^{\I p_z z}}{\sqrt{2\pi}}\frac{\E^{\I m \phi}}{\sqrt{2\pi}} \sqrt{|qB|} \left(\begin{array}{c}
	\tilde{C}^{\bar\sigma}_1 I^{\bar\sigma}_1(\rho) \E^{-\I\phi/2}\\\I C^{\bar\sigma}_2 I^{\bar\sigma}_2(\rho)\E^{+\I\phi/2} \\
	\tilde{C}^{\bar\sigma}_3 I^{\bar\sigma}_1(\rho)\E^{-\I\phi/2} \\ \I C^{\bar\sigma}_4 I^{\bar\sigma}_2(\rho)\E^{\I\phi/2}
\end{array}\right),\quad n\geq 0,\; a \geq 0,
\end{equation}
which are in the same form as in \cite{Sokolov:1986nk}.
The corresponding energy spectrum is
\begin{equation*}
    E = \epsilon\sqrt{p_z^2 + 2n |qB|+M^2}+m\Omega.
\end{equation*}

It can be shown comparing Eq.~(\ref{eq:FInf}) with (\ref{eq:DefHyperF}) that in the infinitely large cylinder the normalization constants satisfy $n N_1^2/N_2^2=1$, taking advantage of that and using the coefficients of transverse polarization
(\ref{eq:CoefficientsTransversePol1Bounded}) and (\ref{eq:CoefficientsTransversePol2Bounded}) we obtain:
\begin{equation}
\label{eq:CoefficientsTransversePolUnb}
\tilde{C}^{\bar\sigma}_{1,3}=-\frac{\bar\sigma}{2\sqrt{2}}B_+(A_+ \pm \zeta A_-),\quad C^{\bar\sigma}_{2,4}=\frac{1}{2\sqrt{2}}B_- (A_- \mp \zeta A_+),
\end{equation}
where
\begin{subequations}
\begin{align}
\label{eq:CoefficientsTransversePolUnb2}
A_\pm = & \left(\frac{E-\Omega\, m \pm p_z}{E-\Omega\, m} \right)^{\frac{1}{2}},\\
B_\pm = & \left(1\pm\frac{\zeta}{\sqrt{(E-\Omega\, m)^2-p_z^2}} \right)^{\frac{1}{2}},
\end{align}
\end{subequations}
in agreement with the solution of  \cite{Sokolov:1986nk}.

It is instructive to study how the unbounded modes are
approached as we increase the radius of the cylinder $R$. For the sake of clarity, we fix
$\bar\sigma=+1$; the same analysis can be done for $\bar\sigma=-1$. In order to have a
wave function that vanish at infinity, we must choose $\lambda$ and $a$ as non-negative
integers. However, in a finite cylinder, solutions of the Dirac equation can also have
modes with
%negative $\lambda$ and
negative $a$. To study what happens to these modes
we evaluate the hypergeometric functions \eq{eq:DefHyperF} for positive and negative integer values of $\lambda=n$ and $a$; we find
\begin{equation}
\label{eq:HyperF1Integer}
\left.\frac{F_1^+(\rho)}{N_1}\right|_{\lambda=n} =\begin{cases}
\sim \E^{-\rho/2} & n \geq 0,\, a\geq 0\\
=0 & n \geq 0,\, a< 0\\
\sim \E^{+\rho/2} & n < 0
\end{cases}
\end{equation}
and
\begin{equation}
\label{eq:HyperF2Integer}
\left.\frac{F_2^+(\rho)}{N_2}\right|_{\lambda=n} =\begin{cases}
%=\E^{+\rho/2} \rho^{-\frac{m+1/2}{2}}\frac{\Gamma(m+1/2)-\Gamma(m+1/2,\rho)}{\Gamma(m+1/2)} \sim \E^{+\rho/2} & n = 0,\, a\geq 0\\
\sim \E^{-\rho/2} & n \geq 1,\, a\geq 0\\
%=\E^{+\rho/2} \rho^{-\frac{m+1/2}{2}} \sim \E^{+\rho/2} & n = 0,\, a< 0\\
=0 & n \geq 1,\, a< 0\\
\sim \E^{+\rho/2} & n \leq 0
\end{cases}
\end{equation}
where $\sim$ denotes the asymptotic behavior at $\rho\to\infty$. The increasing modes, namely those that do not vanish at infinity, are automatically discarded in the unbounded solutions and the spectrum is limited to the case $n\geq 0$ and $a\geq 0$. Moreover, the radial functions are always increasing at infinity when $\lambda$ is not an integer \cite{NISTHandbook}:
\begin{equation}
\frac{F_1(\rho)}{N_1}\sim \E^{+\rho/2},\quad
\frac{F_2(\rho)}{N_2}\sim \E^{+\rho/2},\quad
\text{for }\lambda\notin\mathbb{Z}.
\end{equation}
In a finite cylinder the increasing modes are generally allowed, and we must check what happens as we increase the size of the cylinder.

As anticipated, as the radius of the cylinder increases, the modes of the unbounded case are recovered. Indeed, for $n\geq 0$ and $a\geq 0$, we have 
\begin{equation}
\begin{split}
\lim_{\rho_R\to \infty} F_1^+ =& F_1^+(\lambda=n) = I_{n,a}(\rho),\quad n\geq 0,\; a \geq 0,\\
\lim_{\rho_R\to \infty} F_2^+ =& F_2^+(\lambda=n) = I_{n-1,a}(\rho),\quad n\geq 1,\; a \geq 0.
\end{split}
\end{equation}
The case $n=0$ is recovered if $F_2^+=0$, indeed $I_{n-1,a}(\rho)$ is vanishing for $n=0$.

Consider now the modes that are present in the finite cylinder but are absent in the unbounded space. As noted, the radial functions corresponding to these mode are increasing exponentially with $\rho$. However, the functions $F_1$ and $F_2$ are normalized to $1$ when integrated over the dimensionless radial coordinate $\rho$. This implies that the constants $N_1$ and $N_2$ corresponding to an increasing radial function must fall to zero very rapidly with $\rho_R$
\begin{equation}
\lim_{\rho_R\to \infty} N_{1,2} = 0.
\end{equation}
Eventually, at a large enough  radius, the radial functions are significantly different from zero only close to the boundary of the cylinder
\begin{equation}
\lim_{\rho_R\to \infty} F^+_{1,2} \simeq \delta(\rho-\rho_R),
\end{equation}
because for any finite $\rho$ the radial function is brought to zero by its constant. To wit, the radius of the cylinder increases, these increasing modes, including $F_2^+$ for $n=0$, are pushed and squeezed close to the boundary of the cylinder.
In the limit of infinite radius they can be considered as vanishing everywhere.

In this section, we showed that by imposing that the radial wave functions $f_s$ are vanishing at $r\to \infty$, requires $\lambda$ to be a non-negative integer $n$ and $f_s$ be proportional to the 
generalized Laguerre polynomials \eq{Lagguerre}. In the next section we impose the boundary condition at finite $r$ and as a result $\lambda$ is not an integer and $f_s$ are not expressed in terms of the Laguerre polynomials.

%*********************************************************************************************************
\section{Boundary condition at finite \texorpdfstring{$r=R$}{radius}}\label{sec:MITBC}
%*********************************************************************************************************
A light signal sent radially from any point in the region $r<R=1/\Omega$ will never cross the light cylinder at $r=R$. Therefore,  the wave functions \eq{eq:SolutionDirac} must be confined to the causally connected region $r\le R$. We recently considered a similar problem in connection to the nonrelativistic spin zero bound state problem in rotating frame \cite{Buzzegoli:2022omv}. There we imposed the Cauchy boundary condition at $r=R$ on the wave function that resulted in a peculiar dependence of the energy spectrum on $\Omega$. Whereas the state of the spin zero system is described by the single wave function, a nonrelativistic spin $\tfrac{1}{2}$ state is a two-component spinor, which cannot satisfy the Cauchy boundary condition. This problem is well known since the early days of QCD when the models of hadron structure were developed: the color confinement requires that the quarks be localized within the hadron of a certain radius. Given two solutions of the Dirac equation $(\psi,\,\Phi)$ the quantum theory is well posed if the Hamiltonian is a self-adjoint operator with respect to the inner product, that is $\langle \psi,\,\h{H}\Phi\rangle = \langle \h{H}\psi,\,\Phi\rangle$. In the considered system inside the finite cylinder the Hamiltonian is self-adjoint if the following condition is satisfied, see for instance~\cite{Ambrus:2015lfr}:
\begin{eqnarray}
\label{eq:Hselfadj}
R\int_0^\infty\!\!\!\D z\int_0^{2\pi}\!\!\!\D\phi
    \,\bar{\psi}(R) \gamma^r \Phi(R) =& 0,\\
\de_r (\bar\psi\Phi)(R)=&- \Theta(R),
\end{eqnarray}
where  $\psi(R)=\psi(z,\,\phi,\, r=R)$, $\Theta(R)<\infty$
and $\gamma^r =\slashed{n}=\hat{n}_\mu \gamma^\mu =\hat{r}_\mu \gamma^\mu$, with $\hat{n}=\hat{r}$
the radial direction in cylindrical coordinates.
A consistent solution is to impose the MIT boundary conditions \cite{Chodos:1974je,Shanker:1983yd} that require vanishing of the outward-pointing component of the fermion current, that is the integrand of Eq.~(\ref{eq:Hselfadj}) is vanishing. The MIT boundary condition in the rotating frame were studied in \cite{BezerradeMello:2008zd,Becattini:2011ev,Ambrus:2015lfr}. Other types of boundary conditions have also been considered \cite{Ambrus:2015lfr,Chen:2017xrj,Sadooghi:2021upd}.
In the semi-classical limit, it is more convenient to use the representation obtained in~\cite{Obukhov:2013zca} performing a Foldy-Wouthuysen transformation on a curved background (including rotation), where in the nonrelativistic limit the spinor components related to the antiparticles are vanishing.

The MIT boundary conditions in a cylinder of radius $R$ constrain the wave function to satisfy \cite{Chodos:1974je,Shanker:1983yd}
\begin{equation}
\label{eq:MIT1}
\I \hat{r}^\mu \gamma_\mu \psi(R)= \I \gamma_r \psi(R) = \alpha \psi(R), \quad \alpha=\pm
\end{equation}
and
\begin{equation}
\de_r \left(\bar\psi \psi\right)(R) < \infty,
\end{equation}
where $\hat{r}$ is the radial direction in cylindrical coordinates. This ensures that Eq.~(\ref{eq:Hselfadj}) is satisfied, hence that
the Hamiltonian is self-adjoint, and that the fluxes of energy, momentum, angular momentum
and charge vanish at the boundary~\cite{Ambrus:2015lfr,Buzzegoli:2023eeo}. According to the Hellinger–Toeplitz theorem \cite{reed1980methods} a self-adjoint Hamiltonian  is bounded from below. As a result the vacuum state is well defined which is essential for obtaining meaningful physical results, see Fig.~\ref{fig:energyLevelsP} and discussion in ~\cite{Ambrus:2014uqa,Ambru2021}. Imposing the condition (\ref{eq:MIT1}) to the general solution of the Dirac equation (\ref{eq:SolutionDiracGeneral}), we obtain
\begin{equation}
\begin{cases}
C_1^{\bar\sigma} F^{\bar\sigma}_1(\rho_R) + \alpha C_4^{\bar\sigma} F^{\bar\sigma}_2(\rho_R) = 0, \\
C_3^{\bar\sigma} F^{\bar\sigma}_1(\rho_R) - \alpha C_2^{\bar\sigma} F^{\bar\sigma}_2(\rho_R) = 0,
\end{cases}
\end{equation}
whence
\begin{equation}
\label{eq:MITConstraint1}
F^{\bar\sigma}_1(\rho_R) = -\alpha \frac{C_4^{\bar\sigma}}{C_1^{\bar\sigma}} F^{\bar\sigma}_2(\rho_R)
	=  \alpha \frac{C_2^{\bar\sigma}}{C_3^{\bar\sigma}} F^{\bar\sigma}_2(\rho_R),
\end{equation}
where  $\rho_R=|qB|R^2/2$.
Clearly, the boundary condition can only be satisfied by states such that $C_2^{\bar\sigma}/C_3^{\bar\sigma}=-C_4^{\bar\sigma}/C_1^{\bar\sigma}$.
In the transverse polarization states \eq{eq:CoefficientsTransversePol1Bounded}, it is straightforward to check that
\begin{equation}
\frac{C_4^{\bar\sigma}}{C_1^{\bar\sigma}} =  - \frac{C_2^{\bar\sigma}}{C_3^{\bar\sigma}}
	= \frac{B_4}{B_3}
    = \frac{N_1}{N_2}\zeta\sqrt{\lambda}\sqrt{\frac{\bar{E}-\zeta M}{\bar{E}+\zeta M}}.
\end{equation}
Then, the MIT boundary conditions applied to the transverse polarization states result in a single equation
\begin{equation}
\label{eq:MITConstraint}
\frac{F^{\bar\sigma}_1(\rho_R)}{N_1} = -\alpha  \zeta \sqrt{\lambda}
    \sqrt{\frac{\bar{E}-\zeta M}{\bar{E}+\zeta M}}
    \frac{F^{\bar\sigma}_2(\rho_R)}{N_2}
= -\alpha  \zeta 
    \sqrt{\frac{M^2+\lambda|qB|-\zeta\sqrt{M^2+2\lambda|qB|}}{|qB|}}
    \frac{F^{\bar\sigma}_2(\rho_R)}{N_2},
\end{equation}
where we remind
\begin{equation}
\bar{E} = \sqrt{(\varepsilon E - \Omega\, m)^2-p_z^2}
    = \sqrt{2\lambda|qB|+M^2} .
\end{equation}

Equation (\ref{eq:MITConstraint}) has an infinite countable number of roots $\{\lambda_i\},\,i=1,\,2,\,3,\,\dots$ such that $\lambda_i<\lambda_{i+1}$.
Solving this equation for $\lambda$ provides the complete set of normal modes (\ref{eq:SolutionDiracGeneral})
\begin{equation}
\label{eq:SolutionDiracRhoR}
\psi^{\rho_R}_{i,\,m,\,p_z,\,\zeta}(t,r,\phi,z)=\E^{-\I\epsilon E t}\frac{\E^{\I p_z z}}{\sqrt{2\pi}}\frac{\E^{\I m \phi}}{\sqrt{2\pi}} \sqrt{|qB|} \left(\begin{array}{c}
	C^{\bar\sigma}_1 F^{\bar\sigma}_1(\rho) \E^{-\I\phi/2}\\\I C^{\bar\sigma}_2 F^{\bar\sigma}_2(\rho)\E^{+\I\phi/2} \\
	C^{\bar\sigma}_3 F^{\bar\sigma}_1(\rho)\E^{-\I\phi/2} \\ \I C^{\bar\sigma}_4 F^{\bar\sigma}_2(\rho)\E^{\I\phi/2}
\end{array}\right)_{\lambda=\lambda_i(m,\zeta)},
\end{equation}
and the energy spectrum for the Dirac particle in the finite cylinder, whose energy levels are
\begin{equation*}
    E_{i,m,p_z} = \sqrt{2\lambda_i|qB|+p_z^2+M^2}+\Omega\,m\,.
\end{equation*}
Fixing the value of $\rho_R$, $M$, $|qB|$, $\bar\sigma$ and $\alpha$ fixes the spectrum. In general the roots of this equation depend on $m$ and $\zeta$.
One can check that when increasing $\rho_R$ the positive roots are
\begin{equation}
\lim_{\rho_R\to\infty} \lambda = \begin{cases}
    0,\, 1,\, 2,\, \dots & \zeta=\alpha \\
    1,\, 2,\, \dots & \zeta=-\alpha
\end{cases}
\end{equation}
independently of $m$ and $\zeta$. In the unbounded case ($R\to \infty$) the lowest Landau level $\lambda=0$ has both positive and negative polarization $\zeta$, while with MIT BC only the polarization $\zeta=\alpha$ admits roots with $\lambda\simeq 0$. In \cite{Buzzegoli:2023eeo} we investigated implications of this feature for the chiral magnetic current in a finite cylinder and we found that despite this difference the results with MIT BC approach the unbounded results as $\rho_R\to \infty$. In this work we set $\alpha=-1$.

%%%%%%%%%%%%%%%%%
\begin{figure}[bth]
    \centering
    \includegraphics[width=.45\textwidth]{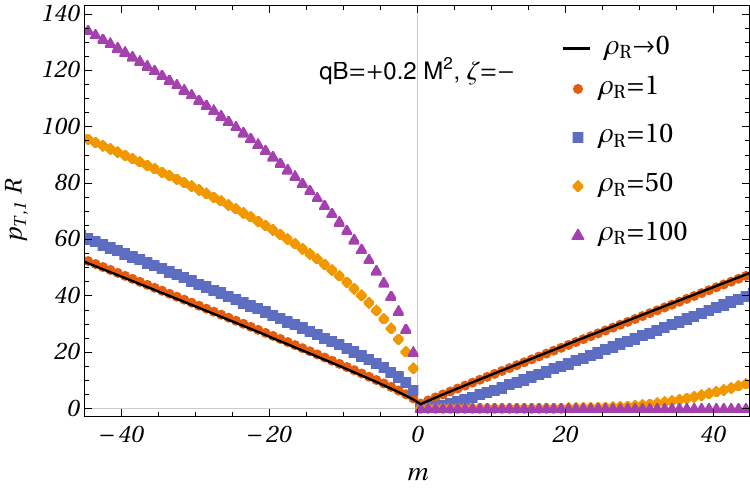}
    \caption{The lowest transverse momentum $p_{T,1}=\sqrt{2|qB|\lambda_1}$ in units of the cylinder radius $R$ as a function of the angular momentum $m$ for various $\rho_R$’s and fixed negative polarization $\zeta=-1$ and magnetic field $qB=+0.2 M^2$.}
    \label{fig:pt}
\end{figure}
%%%%%%%%%%%%%%%
 
The lowest transverse momentum $p_{T,1}=\sqrt{2|qB|\lambda_1(m,\zeta)}$ resulting from MIT BC is shown in Fig.~\ref{fig:pt} for a fixed value of magnetic field. The solutions with $m<-1/2$ are not allowed in the unbounded case, as they would imply a negative radial quantum number (\ref{def.a}), instead they are possible in a finite volume.
Note that for larger system size, that is for larger $\rho_R$, the first root is $\lambda_1\simeq 0$ up to larger values of $m$ as expected for a cylinder approaching the infinite volume limit.
The lowest transverse momentum resulting from MIT BC shown in Fig.~\ref{fig:pt} has the same qualitative features as that obtained in \cite{Chen:2017xrj} with a similar boundary condition. The main difference is that the BC used in \cite{Chen:2017xrj} depends only on $\rho_R$, while MIT BC depends on $\zeta$, $M$ and $qB$ as well (hence also on $R$ through $\rho_R$). Despite being more complicated, we nevertheless used the MIT BC to make sure that all the physical quantities remain confined in the cylinder. Indeed, MIT BC implies that the flux of the current, energy, momentum and angular momentum are vanishing at the cylinder surface~\cite{Buzzegoli:2023eeo}.

\subsection{Energy levels}
%%%%%%%%%%%%%%%%%
\begin{figure}[th]
    \centering
    \includegraphics[width=.45\textwidth]{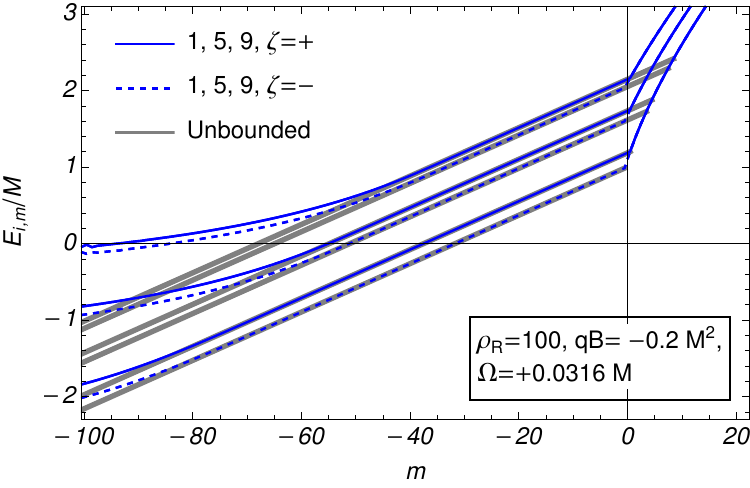}
    \includegraphics[width=.45\textwidth]{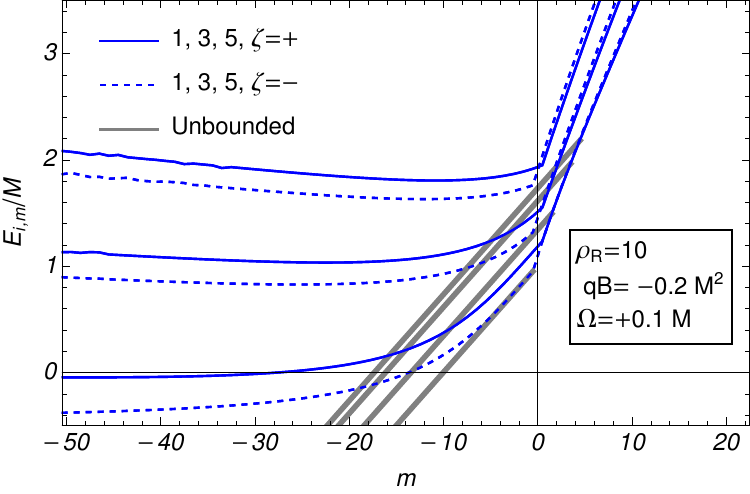}\\
    \includegraphics[width=.45\textwidth]{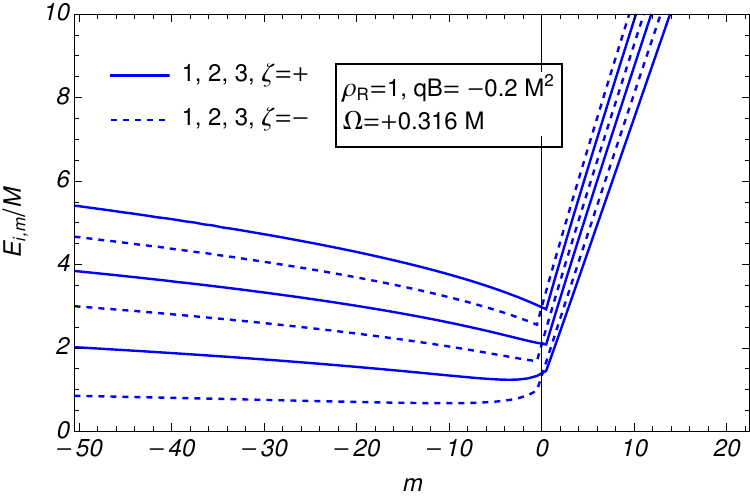}
    \includegraphics[width=.45\textwidth]{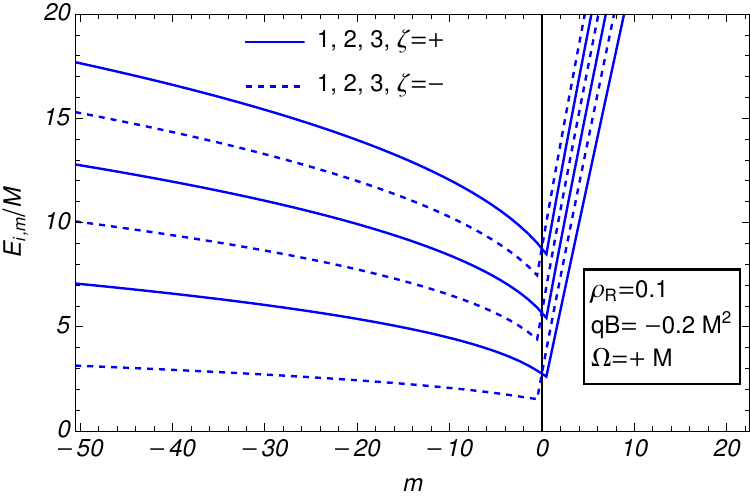}\\
    \caption{The energy levels $E_{i,m,\zeta}(qB,\rho_R)$ as a function of $m$ with $\bar\sigma=-1$. The index $i=1,\,2,\,3,\dots$ denotes the first, second and third root $\{\lambda_i\}$ of the MIT boundary condition. Solid lines correspond to $\zeta=+$, dashed one to $\zeta=-$. Blue lines: finite $\rho_R$ as indicated, gray lines: $\rho_R\to \infty$ (unbounded volume). }
    \label{fig:energyLevelsP}
\end{figure}
%%%%%%%%%%%%%%%

We are interested in the case where external magnetic field and rotation are both present. As argued, we choose to confine the system in a cylinder of radius $R=1/\Omega$.
In this scenario the energy levels are given by
\begin{equation}
    E_{i,\,m,\,\zeta,\,p_z} = \sqrt{p_z^2 + 2\lambda_i(m,\zeta) |qB|+M^2}+m\Omega.
\end{equation}

Figure~\ref{fig:energyLevelsP} shows some of the energy levels at fixed magnetic field $qB=-0.2 M^2$ with $\bar\sigma=-1$ and at $p_z=0$ for various values of positive angular velocity, that is directed along the $z$ axis. For the smallest angular velocities we compare the energies inside a cylinder with the results obtained in the unbounded case. As noted earlier, at large $\rho_R$ the energy levels from MIT and unbounded solution coincide up to some value of $m$. In the unbounded case the energy levels keep decreasing to $-\infty$ as we decrease $m$. Instead, with MIT BC the energy levels reach a minimum and then the energy increases. Differently from the Landau levels without rotation, in the presence of rotation the energy can be smaller than $M$ and can also be negative because the potential energy associated with rotation can be negative. For higher values of angular velocity the gaps between levels is larger because the radius of Landau orbits is comparable with the radius of the cylinder and quantum effects are predominant. Energies are also positive and large because it requires more energy to keep the fermion inside the volume, as follows from the uncertainty principle. At large angular velocities we are not comparing the results with the unbounded solution because the latter break causality.

%%%%%%%%%%%%%%%%%
\begin{figure}[th]
    \centering
    \includegraphics[width=.45\textwidth]{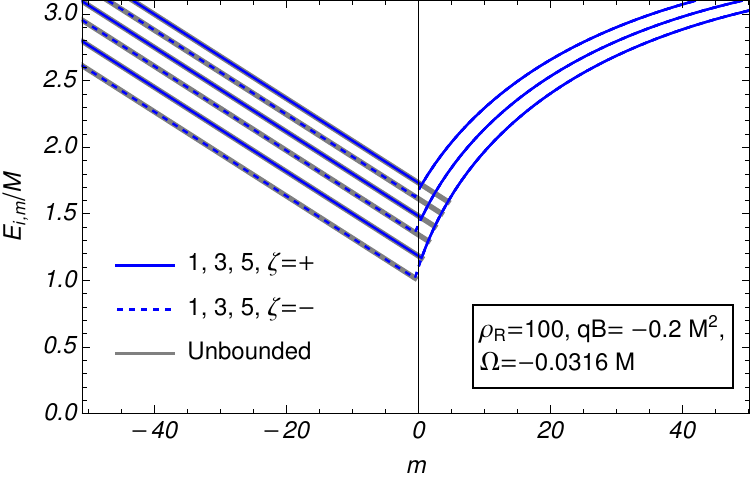}
    \includegraphics[width=.45\textwidth]{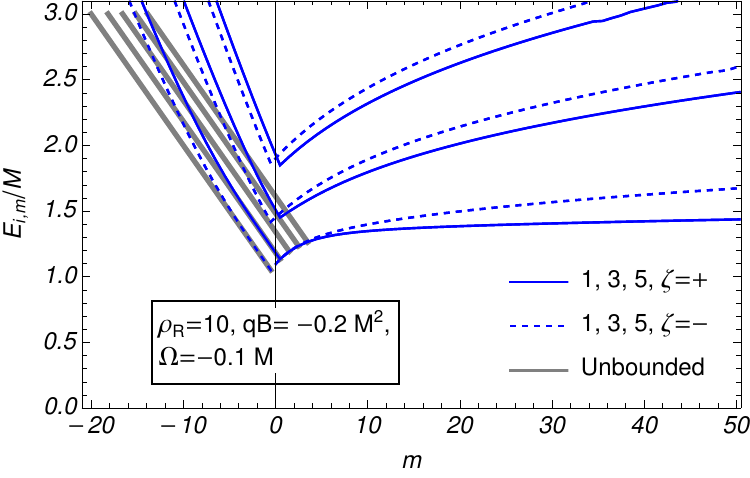}\\
    \includegraphics[width=.45\textwidth]{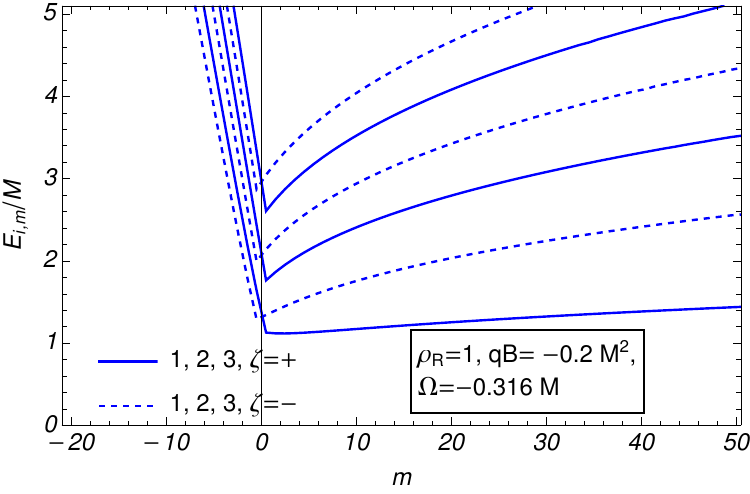}
    \includegraphics[width=.45\textwidth]{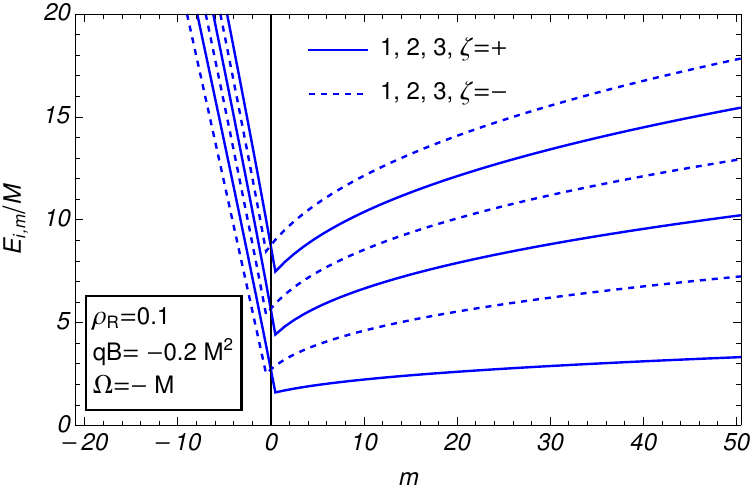}\\
    \caption{Same as Fig. \ref{fig:energyLevelsP} but for opposite sense of rotation.}
    \label{fig:energyLevelsM}
\end{figure}
%%%%%%%%%%%%%%%

The energy levels of a system rotating in the opposite direction (``negative rotation") are shown in Fig.~\ref{fig:energyLevelsM}. The properties of a rotating fermion in magnetic field and its radiation under the change of charge, magnetic field and rotation direction were analyzed in~\cite{Buzzegoli:2023vne}. Similarly to the positive rotation, at small angular velocities the MIT and unbounded solution results coincide except for the values of $m$ that are not allowed in the unbounded case. It might seem from the figure that in this case the energy levels resulting from the unbounded solution remain positive. However, the energy is $E_{n,m}=\sqrt{2n|qB+M^2}+m\Omega$ and for very large and negative $m$ the energy is still going to $-\infty$. By comparing Figs.~\ref{fig:energyLevelsP} and \ref{fig:energyLevelsM} at $\rho_R=100$ it is clear that the number of states up to a given energy is much smaller in the negative rotation case. For this reason, in the following sections, the radiation for positive rotation and large value of $\rho_R$ have been evaluated to lower values of energy. Note that this is not a problem in the unbounded case, because the analytical solutions $\lambda=n$ do not have to be computed numerically for every value of $m$, $\zeta$, $\rho_R$ and $qB$.

%%%%%%%%%%%%%%%%%
\begin{figure}[th]
    \centering
    \includegraphics[width=.45\textwidth]{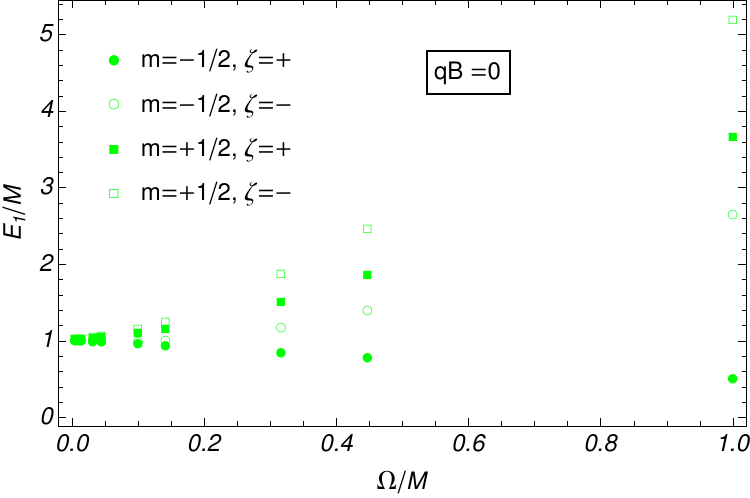}
    \includegraphics[width=.45\textwidth]{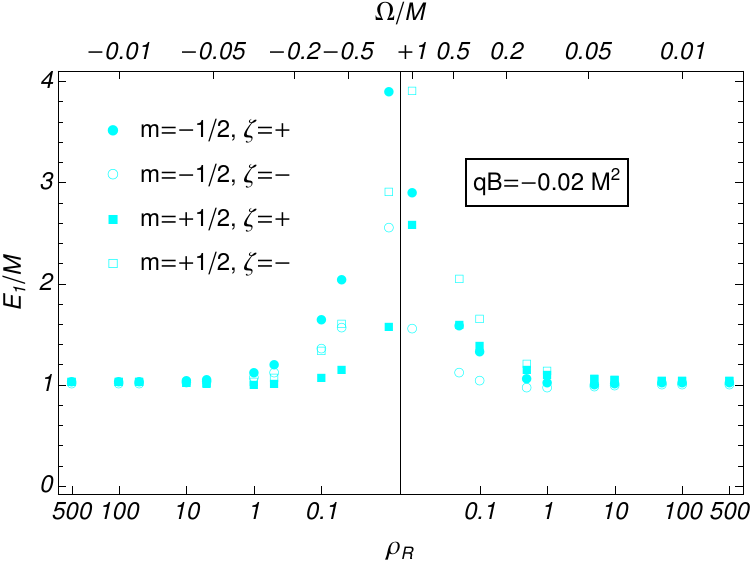}\\
    \includegraphics[width=.45\textwidth]{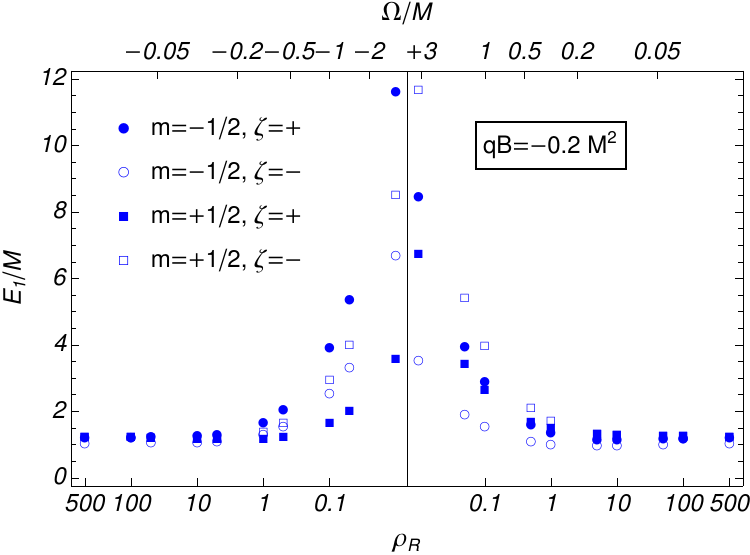}
    \includegraphics[width=.45\textwidth]{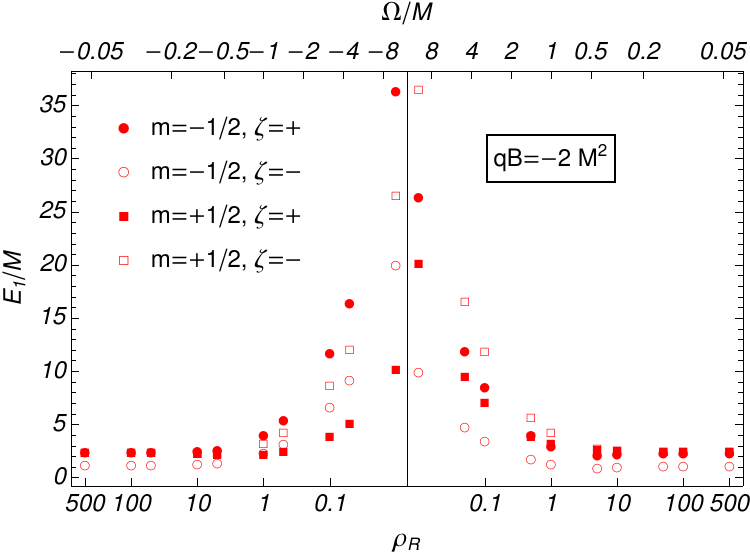}\\
    \caption{The first energy level $E_1=E_{i=1,m,\zeta}(qB,\rho_R)$ as a function of $\rho_R$ for $\zeta=\pm$, $m=\pm 1/2$ and the following values of $qB/M^2=0,\,-0.02,\,-0.2$ and $-2$. In the top left panel $qB=0$ and since $\rho_R=|qB|/(2\Omega^2)$ is always zero we only indicate the value of angular velocity on the horizontal axis. For finite value of $qB$ (other three panels) we indicate the values of $\Omega/M$ on the top and the corresponding $\rho_R$ on the bottom of each panel. }
    \label{fig:energy_vs_RhoR}
\end{figure}
%%%%%%%%%%%%%%%

We study the dependence of the energy levels on $qB$  in Fig.~\ref{fig:energy_vs_RhoR} where we show the first energy level $E_1$, that is the energy corresponding to the first root $\lambda_1$, with $m=1/2$ and $m=-1/2$ and both polarizations $\zeta=\pm$ at different values of $qB$ as a function of $\Omega$, or equivalently of $\rho_R$.
As expected, the energy increases with $|qB|$ and $|\Omega|$, and approaches the usual Landau levels as the angular velocity decreases. In the absence of the magnetic field, $qB=0$, whose solutions are given below, the energy is lower when the angular momentum is in the opposite direction of rotation ($m=-1/2$ in the figure) according to the energy shift $\vec{\Omega}\cdot\vec{J}$. For $qB=0$, by symmetry, flipping the direction of rotation is the same as flipping the sign of $m$; the physics of the problem does not change, only the reference system has been reflected. At finite magnetic field when rotation and $q\vec{B}$ are antialigned (e.g.\ $qB<0$ and $\Omega >0$) the lowest level is obtained when angular momentum is antiparallel to the rotation, and hence parallel to $q\vec{B}$. When rotation and $q\vec{B}$ are aligned (e.g.\ $qB<0$ and $\Omega <0$) the lowest level is obtained when angular momentum is antiparallel to the rotation for $\rho_R<1$ (rapid rotation), and parallel to $q\vec{B}$ for $\rho_R>1$ (slow rotation). To understand what happens when $qB$ is positive one has simply to read the results for negative $qB$ and change the sense of rotation, as we explained in detail in \cite{Buzzegoli:2023vne}.

%---------------------------------------------------------------------
\section{Vanishing magnetic field}
\label{sec:JustRotation}
%---------------------------------------------------------------------
For fast rotation and small $\rho_R$, the effect of the magnetic field is a small correction to the predominant effect of rotation. It is instructive to compare this case with the one with a vanishing magnetic field. Indeed, a charged fermion embedded in a rotating system is accelerating because of its rotation and it will radiate even without the need of an external magnetic field. We compute the radiation of such a fermion in a finite cylinder with radius $R=1/\Omega$ assuming MIT BC.

To obtain the complete set of normal modes and the energy spectrum we follow the same procedure described above with the Hamiltonian $H$ given by Eq. (\ref{eq:H}) with vanishing gauge field. The results are the same except that the radial functions $F_{1,2}$ are now given by Bessel functions and the transverse momentum $p_T$ is not related to the magnetic field: in the limit of vanishing magnetic field $2\lambda|qB|\to\, p_T$. The general form of solutions to the Dirac equation with just rotation (JR), viz.\  in rotating frame and vanishing magnetic field, is \cite{Mameda-thesis}
\begin{equation}
\label{eq:SolutionDiracGeneralJR}
\psi^{\rm JR}(t,r,\phi,z)=\E^{-\I\epsilon E t}\frac{\E^{\I p_z z}}{\sqrt{2\pi}}\frac{\E^{\I m \phi}}{\sqrt{2\pi}}  \left(\begin{array}{c}
	C^{\rm JR}_1 F^{\rm JR}_1(r) \E^{-\I\phi/2}\\
	(-\I) C_2^{\rm JR} F^{\rm JR}_2(r)\E^{+\I\phi/2} \\
	C^{\rm JR}_3 F^{\rm JR}_1(r)\E^{-\I\phi/2} \\
	(-\I) C^{\rm JR}_4 F^{\rm JR}_2(r)\E^{\I\phi/2}
\end{array}\right),
\end{equation}
with $F^{\rm JR}_{1,2}$ given by
\begin{equation}
\label{eq:RadialFJR}
\begin{split}
F^{\rm JR}_1(r) = & N^{\rm JR}_{m-1/2} J_{m-1/2}(p_T R),\\
F^{\rm JR}_2(r) = & N^{\rm JR}_{m+1/2} J_{m+1/2}(p_T R),
\end{split}
\end{equation}
with $J$ the Bessel function and with normalization constant
\begin{equation}
N^{\rm JR}_\nu = \sqrt{2}\Omega \left|J_\nu\left(\frac{p_T}{\Omega}\right)^2
    +J_{\nu+1}\left(\frac{p_T}{\Omega}\right)^2
    -\frac{2\nu\Omega}{p_T}J_{\nu}\left(\frac{p_T}{\Omega}\right)J_{\nu+1}\left(\frac{p_T}{\Omega}\right)\right|^{-1/2},
\end{equation}
such that
\begin{equation}
\int_0^{1/\Omega}\D r\, r \left| F^{\rm JR}_{1,2}(r)\right|^2 = 1.
\end{equation}
In the transverse polarization the coefficients $C^{\rm JR}_s$ read
\begin{equation}
\label{eq:CoefficientsTransversePol1BoundedJR}
C^{\rm JR}_{1,3}=\frac{1}{2\sqrt{2}}B^{\rm JR}_+(A^{\rm JR}_+ \pm \zeta A^{\rm JR}_-),\quad
C^{\rm JR}_{2,4}=\frac{1}{2\sqrt{2}} \frac{N^{\rm JR}_{m-1/2}}{N^{\rm JR}_{m+1/2}}
    B^{\rm JR}_- (A^{\rm JR}_- \mp \zeta A^{\rm JR}_+),
\end{equation}
where 
\begin{equation}
\begin{split}
\label{eq:CoefficientsTransversePol2BoundedJR}
A^{\rm JR}_\pm = & \left(\frac{E-\Omega\, m \pm p_z}{E-\Omega\, m} \right)^{\frac{1}{2}},\\
B^{\rm JR}_\pm = & \left(\frac{2(\bar{E}\pm\zeta M)}{\left(1+\left(\frac{N^{\rm JR}_{m-1/2}}{N^{\rm JR}_{m+1/2}}\right)^2\right)\bar{E} + \left(1-\left(\frac{N^{\rm JR}_{m-1/2}}{N^{\rm JR}_{m+1/2}}\right)^2\right)\zeta M} \right)^{\frac{1}{2}},
\end{split}
\end{equation}
with $\bar{E} = \sqrt{(\varepsilon E - \Omega\, m)^2-p_z^2} = \sqrt{p_T^2+M^2}$.

Likewise, the MIT boundary condition is imposed in the same fashion as described for the case with magnetic field. At the end we obtain that the MIT BC impose a constrain expressed as in Eq.~(\ref{eq:MITConstraint1}) with the hypergeometric function replaced with the Bessel functions; the constrain reads:
\begin{equation}
\label{eq:MITConstraintJR}
J_{m-1/2}\left(\frac{p_T}{\Omega}\right) - \alpha\zeta \sqrt{\frac{\sqrt{p_T^2 +M^2}-\zeta M}{\sqrt{p_T^2 +M^2}+\zeta M}}
J_{m+1/2}\left(\frac{p_T}{\Omega}\right) = 0 .
\end{equation}
Solving this equation for the roots $\{p_{T,i}\}$ provides the complete set of normal modes and the energy spectrum for the rotating Dirac particle in the finite cylinder.
As in the finite magnetic field case, the roots $p_T$ depend on $\Omega$, $m$ and $\zeta$.

%%%%%%%%%%%%%%%%%
\begin{figure}[th]
    \centering
    \includegraphics[width=.45\textwidth]{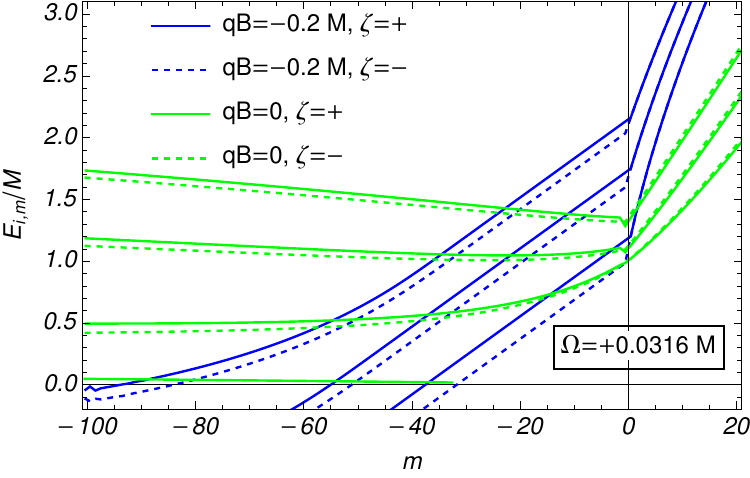}
    \includegraphics[width=.45\textwidth]{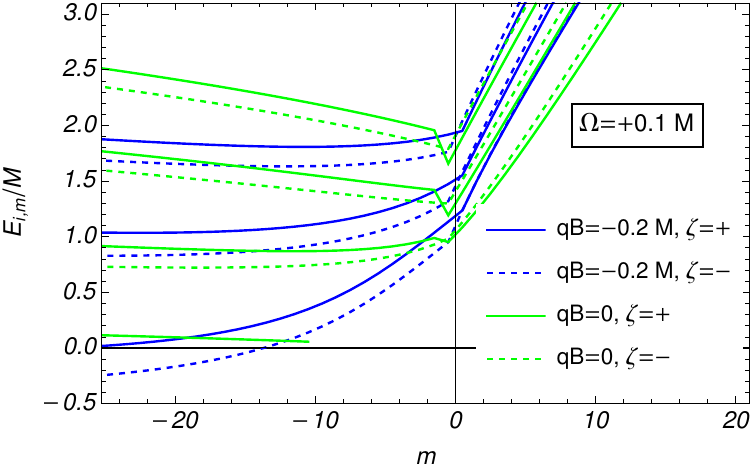}\\
    \includegraphics[width=.45\textwidth]{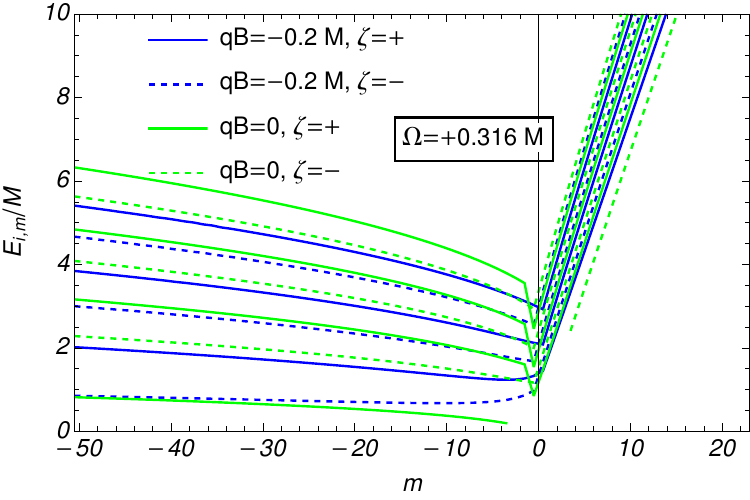}
    \includegraphics[width=.45\textwidth]{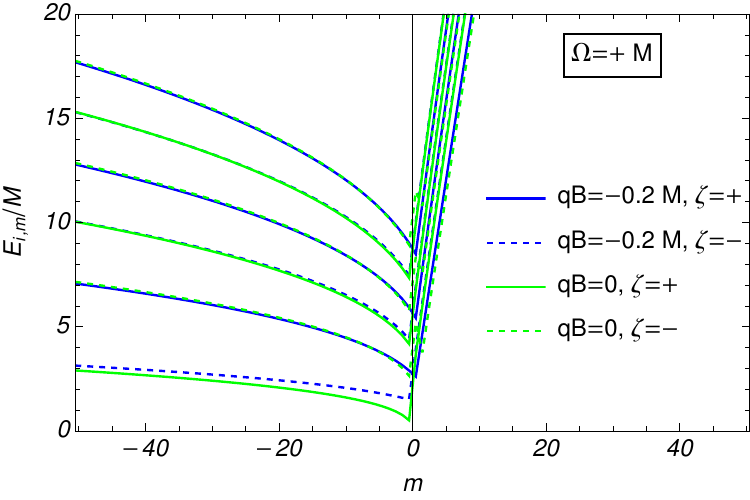}
    \caption{Green lines: the energy levels $E_{i,m,\zeta}(\Omega)$ at $B=0$ as a function of $m$ for various values of $\Omega$. The index $i=1,\,2,\,3,\dots$ denotes the first, second and third root $\{p_{T,i}\}$ of the MIT boundary condition. Solid lines correspond to $\zeta=+$, dashed ones to $\zeta=-$. Blue lines: the levels at finite magnetic field shown in Fig.~\ref{fig:energyLevelsP}. }
    \label{fig:energyLevelsJR}
\end{figure}
%%%%%%%%%%%%%%%
%%%%%%%%%%%%%%%%%
\begin{figure}[th]
    \centering
    \includegraphics[width=.45\textwidth]{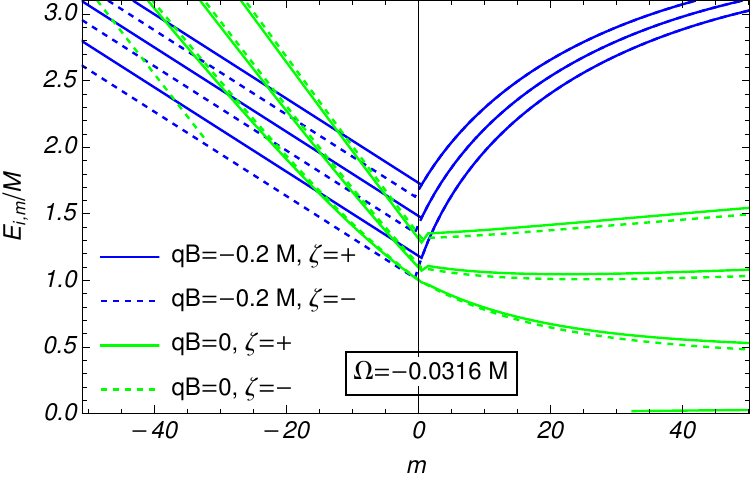}
    \includegraphics[width=.45\textwidth]{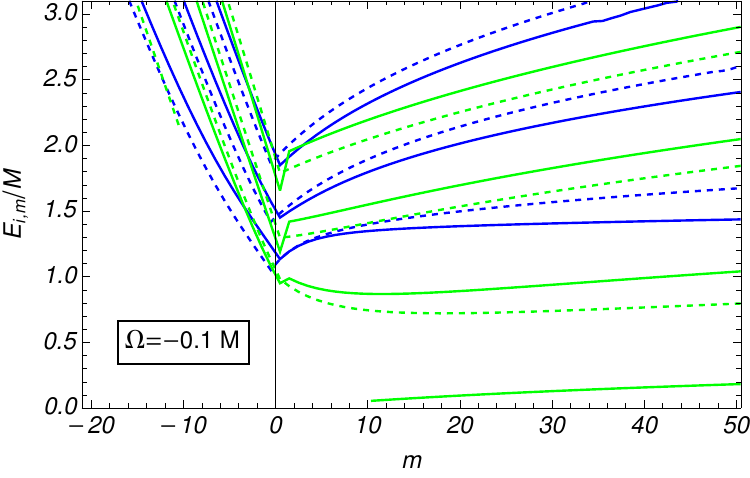}\\
    \includegraphics[width=.45\textwidth]{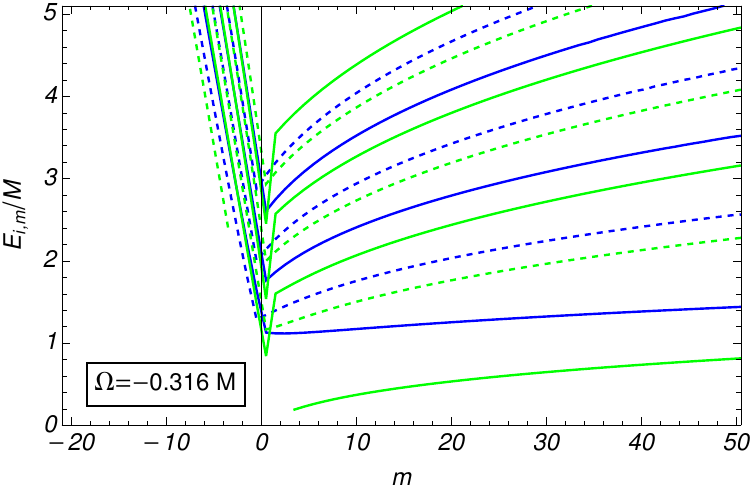}
    \includegraphics[width=.45\textwidth]{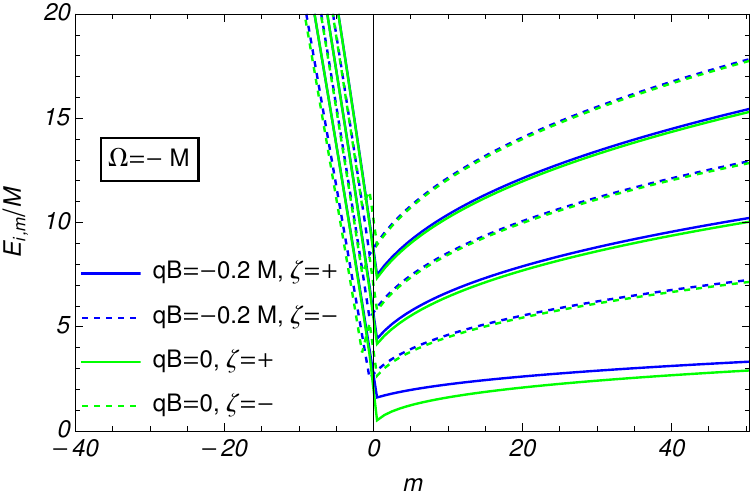}
    \caption{Same as Fig.~\ref{fig:energyLevelsJR} but for negative direction of rotation. }
        \label{fig:energyLevelsJRN}
\end{figure}
%%%%%%%%%%%%%%%
The energy levels
\begin{equation}
    E_{i,\,m,\,\zeta,\,p_z} = \sqrt{p_z^2 + p_{T,i}(\zeta,\,m)^2 +M^2}+m\Omega
\end{equation}
at zero longitudinal momentum $p_z=0$ are shown in Fig.~\ref{fig:energyLevelsJR} and~\ref{fig:energyLevelsJRN} for different values of angular velocity (green lines) in comparison with the previous one obtained at finite magnetic field (blue lines). In Fig.~\ref{fig:energyLevelsJR} the energy levels at zero magnetic field are compared to the case when a finite $q\vec{B}$ is antiparallel to the rotation and in Fig.~\ref{fig:energyLevelsJRN} when $q\vec{B}$ is parallel to $\vec{\Omega}$.
As expected the energy spectra with or without magnetic field are similar for large values of the angular velocity. Moreover, we note that the MIT BC with just rotation seems to discard negative energies. As we change $m$ new roots with lower values of $p_T$ becomes possible whenever the corresponding energy turns positive.

Now that we have the complete set of solutions of a rotating charged fermion in the causal cylinder we can move on to discuss the electromagnetic radiation due to  transitions between the fermion energy levels.

%---------------------------------------------------------------------
\section{Differential Radiation Intensity}
\label{sec:RadiationInt}
%---------------------------------------------------------------------

Thus far we investigated the fermion spectrum in rotating frame subject to the external magnetic field. We observed that it significantly differs from the familiar Landau levels. Thus its electromagnetic radiation must be quite distinct from the synchrotron radiation as indeed we found in  \cite{Buzzegoli:2022dhw,Buzzegoli:2023vne}. It is convenient to study the electromagnetic radiation in the stationary laboratory frame where the fermion motion is a nonlinear combination of circular motion in the external magnetic field and another circular motion with angular velocity $\Omega$ as the fermion is dragged along with the rotating medium. Unlike the fermion, the electromagnetic field does not rotate. Therefore, in the laboratory frame the photon wave function is given by the standard inertial frame solutions to the wave equation. This setup is described in detail in \cite{Buzzegoli:2023vne} where it is also elucidated using a classical model.

The intensity of electromagnetic radiation by a fermion of charge $q$ in magnetic field $\b B=B\unit z$, rotating together with a larger system with angular velocity $\b\Omega=\Omega \unit z$ in the unbounded volume is given by \cite{Buzzegoli:2022dhw}
\begin{equation}
\label{eq:RadIntensityUnb}
W_{n,a,p_z,\zeta} = \frac{q^2}{4\pi}  \sum_{n',a',\zeta'}\sum_{l,h}\delta_{m',m-l}\int  \left|
\mean{\,\vec{j}\cdot\vec{\Phi}\,}
\right|^2 \delta(\omega-E+E')\omega^2 \sin\theta d\omega d\theta\,,
\end{equation}
where $\theta$ is the angle between the photon momentum and the rotation axis and the matrix elements $\mean{\,\vec{j}\cdot\vec{\Phi}\,}$ are defined as
\begin{align}
\label{eq:MatrixElementsTPUnb}
 \int \bar\psi^\infty_{n',\,a',\,p_z',\,\zeta'}(\b x) \bm \Phi^*_{h,l,\omega}(\b x)\cdot \bm \gamma \psi^\infty_{n,\,a,\,p_z,\,\zeta}(\b x)\, d^3x \equiv \mean{\,\vec{j}\cdot\vec{\Phi}\,}\delta_{m',m-l}\frac{ 2\pi}{L} \delta(p_z-p_z'-k_z)\, ,
\end{align}
with the fermion wave functions given by Eq. (\ref{eq:SolutionDirac}).
The primed quantities throughout the present and the  remaining sections  refer to the Landau levels of the final state unless indicated otherwise, $L$ is the size of the system along the rotation axis.

 It is straightforward to generalize 
\eq{eq:RadIntensityUnb}, \eq{eq:MatrixElementsTPUnb}
 to the case of the finite cylinder. To this end we replace the wave functions (\ref{eq:SolutionDirac}) with the ones satisfying the MIT BC given in Eq. (\ref{eq:SolutionDiracRhoR}) and the sums over the quantum numbers $\{n,\,a,\,\zeta,\p_z\}$ with the sums over $\{i,\,m,\,\zeta,\p_z\}$. Notice that since the principal quantum number $\lambda_i(m,\,\zeta)$ depends on $m$ and $\zeta$, in the finite volume we necessarily have to sum over the roots $i$ specifying the values of $m$ and $\zeta$. Moreover, the  integral in  (\ref{eq:MatrixElementsTPUnb}) is now restricted to the volume of the cylinder $V(R)$.
Explicitly, we are now computing the intensity of electromagnetic radiation
\begin{equation}
\label{eq:RadIntensity}
W_{i,\,m,\,p_z,\,\zeta} = \frac{q^2}{4\pi}  \sum_{i',\,m',\,\zeta'}\sum_{l,h}\delta_{m',m-l}\int  \left|
\mean{\,\vec{j}\cdot\vec{\Phi}\,}
\right|^2 \delta(\omega-E+E')\omega^2 \sin\theta d\omega d\theta\,,
\end{equation}
where
\begin{align}
\label{eq:MatrixElementsTP}
 \int_{V(R)} \bar\psi^{\rho_R}_{i',\,m',\,p_z',\,\zeta'}(\b x) \bm \Phi^*_{h,l,\omega}(\b x)\cdot \bm \gamma \psi^{\rho_R}_{i,\,m,\,p_z,\,\zeta}(\b x)\, d^3x \equiv \mean{\,\vec{j}\cdot\vec{\Phi}\,}\delta_{m',m-l}\frac{ 2\pi}{L} \delta(p_z-p_z'-k_z)\, .
\end{align}

We assume that the emitted photon does not interact with the rotating system and can escape the finite cylinder without interactions, therefore its wave function is independent of $\Omega$ in the laboratory (stationary) frame. The wave function of photon with energy $\omega$, angular momentum projection on the axis of rotation $l=0,\pm 1, \pm 2, \ldots$ and helicity $h=\pm 1$ reads (see e.g.\  \cite{Buzzegoli:2023vne}):
\begin{equation}
\label{eq:TwistedPhotonWF}
\b\Phi=\frac{\omega}{k_\perp} \frac{1}{\sqrt{2}}
    \left(h\,\b T + \b P \right)e^{i\left(k_{z} z +l \phi\right)}\equiv \b\varphi e^{i\left(k_{z} z +l \phi\right)}\,,
\end{equation}
where the toroidal and poloidal functions read
\begin{align}
\b T&=\frac{i l}{k r} J_l\left(k_{\perp} r\right) \unit{r}-\frac{k_{\perp}}{k} J_{l}^{\prime}\left(k_{\perp} r\right) \unit{\phi}\label{eq:Toroidal}\\
\b P&=\frac{i k_z k_\perp}{k^2} J_l^{\prime}\left(k_{\perp} r\right) \unit{r}-\frac{l k_{z}}{k^2 r} J_{l}\left(k_{\perp} r\right) \unit{\phi} + \frac{k_\perp^2}{k^2} J_{l}\left(k_{\perp} r\right) \unit{z} .%e^{i\left(k_{z} z +l \phi\right)}\,.
\label{eq:Poloidal}
\end{align}
In this formula $J'_l(z)=dJ_l(z)/dz$.
Using the identity
\begin{equation}\label{c10}
J_l'(z)=\frac{1}{2}\left(J_{l-1}(z)
    -J_{l+1}(z)\right),
\end{equation}
we obtain the components of the photon wave function in cylindrical coordinates
\begin{subequations}\label{phivector}
\begin{align}
\varphi^*_r(r) = & \I \frac{k_z}{k}\frac{J_{l+1}(k_\perp r)
    -J_{l-1}(k_\perp r)}{2}
    -\frac{\I h l}{k_\perp r}J_l(k_\perp r),\\
\varphi^*_\phi(r) = & \frac{h}{2}J_{l+1}(k_\perp r)
    -\frac{h}{2}J_{l-1}(k_\perp r)
    -\frac{l k_z}{k k_\perp r} J_l(k_\perp r),\\
\varphi^*_z(r) = & \frac{k_\perp}{k} J_l(k_\perp r).
\end{align}
\end{subequations}
To obtain the Cartesian components of the fermion current we use the explicit form of the wave functions (\ref{eq:SolutionDiracGeneral}) (and suppress the $\bar\sigma$ superscript throughout this section): 
\begin{subequations}\label{MatrElCylinder}
\begin{align}
f_{\lambda',a'}^{T} \gamma^0 \gamma^x f_{\lambda,a} =& \I |q B|\left[
    K_1\E^{\I\phi}F_2 F_1' - K_2\E^{-\I\phi}F_1 F_2'\right]\nonumber\\
    \equiv & |q B| (\I F_x^+ \E^{\I\phi} - \I F_x^- \E^{-\I\phi}),\\
f^T_{\lambda',a'} \gamma^0 \gamma^y f_{\lambda,a} =& |q B|\left[
    K_1\E^{\I\phi}F_2 F_1' +K_2\E^{-\I\phi}F_1 F_2'\right]\nonumber\\
    \equiv & |q B| ( F_x^+ \E^{\I\phi} + F_x^- \E^{-\I\phi}),\\
f^T_{\lambda',a'} \gamma^0 \gamma^z f_{\lambda,a} =& |q B|\left[
    K_4 F_1 F_1' -K_3 F_2 F_2'\right]
    \equiv |q B| F_z,
\end{align}
\end{subequations}
where we used the definitions
\begin{equation}
\label{eq:ABCD}
\begin{split}
    K_1 =& C_1' C_4 + C_3' C_2\,,\quad
    K_2 = C_4'C_1 + C_2' C_3,\\
    K_3 =& C_4' C_2 + C_2' C_4 \,,\quad
    K_4 = C_1'C_3 + C_3' C_1\, .
\end{split}
\end{equation}
Writing the scalar product $\vec{\varphi}^*\cdot\vec{\gamma}$ in the form
\begin{equation*}
\vec{\varphi}^*\cdot\vec{\gamma} =
    \left(\cos\phi \varphi_r^* - \sin\phi\varphi_\phi^* \right)\gamma^x
    +\left(\sin\phi \varphi_r^* + \cos\phi\varphi_\phi^*\right)\gamma^y
    + \varphi_z^* \gamma^z
\end{equation*}
and using (\ref{phivector}) and (\ref{MatrElCylinder}) the matrix element (\ref{eq:MatrixElementsTP}) becomes
\begin{align}\label{AverJPHiCylinder}
\mean{\,\vec{j}\cdot\vec{\Phi}\,}=  |qB|
   \int_0^R dr r &
    \left[K_1 F_2 F_1'\left(\I \varphi_r^*(r) + \varphi_\phi^*(r)\right) - K_2 F_1 F_2'\left(\I \varphi_r^*(r) - \varphi_\phi^*(r)\right)\right.\nonumber\\ & \left.
        +(K_4 F_1 F_1' - K_3 F_2 F_2' )\varphi_z^*(r)\right]\,.
\end{align}
Using another recurrence relation of the Bessel functions
\begin{equation*}
    \frac{2\nu}{z} J_{\nu}(z) = J_{\nu+1}(z) + J_{\nu-1}(z)\,
\end{equation*}
we can reduce (\ref{AverJPHiCylinder}) to the following form
\begin{align}\label{AverPHI3Cylinder}
\mean{\,\vec{j}\cdot\vec{\Phi}\,}=
\sin\theta\left[K_4\mathcal{J}_4 - K_3 \mathcal{J}_3\right]
 + K_1\left(h-\cos\theta\right)\mathcal{J}_1 - K_2\left(h+\cos\theta\right)\mathcal{J}_2\, 
\end{align}
where we defined 
\begin{equation}
\label{eq:RadialIntzCylinder}
\begin{split}
\mathcal{J}_{1}(x;\rho_R)&=\int_0^{\rho_R} J_{m-m'+1}\left(2(x\rho)^{1/2} \right) F_{2}(\rho) F_{1}'(\rho) \D\rho\,,\\
\mathcal{J}_{2}(x;\rho_R)&=\int_0^{\rho_R} J_{m-m'-1}\left(2(x\rho)^{1/2} \right) F_{1}(\rho) F_{2}'(\rho) \D\rho\,,\\
\mathcal{J}_{3}(x;\rho_R)&=\int_0^{\rho_R} J_{m-m'}\left(2(x\rho)^{1/2} \right) F_{2}(\rho) F_{2}'(\rho)\D\rho \,,\\
\mathcal{J}_{4}(x;\rho_R)&=\int_0^{\rho_R} J_{m-m'}\left(2(x\rho)^{1/2} \right) F_{1}(\rho) F_{1}'(\rho)\D\rho \,,
\end{split}
\end{equation}
and $x=k_\bot^2/(2|qB|)$ is a dimensionless variable. 
Inserting  (\ref{AverPHI3Cylinder}) into (\ref{eq:RadIntensity})   yields the expression for the differential radiation intensity for a photon with the circular polarization $h$:
\begin{align}\label{eq:DiffIntensityCylinder}
   W_{i,m,p_z,\zeta}^{h} = &\frac{q^2}{4\pi}  \sum_{i',m',\zeta'}\int \omega^2\sin\theta d\omega\, d\theta\, \delta(\omega-E+E') \nonumber\\
&\times \Big|
\sin\theta\left[K_4\mathcal{J}_4 - K_3 \mathcal{J}_3\right] + K_1\left(h-\cos\theta\right)\mathcal{J}_1 - K_2\left(h+\cos\theta\right)\mathcal{J}_2
\Big|^2\,. 
\end{align}

The total radiation intensity requires integration over $\omega$ which is easily done by the virtue of the delta function in (\ref{eq:DiffIntensityCylinder}). The result is simplest in the reference frame where the initial fermion has $p_z=0$. We restrict our calculation to that frame henceforth. The energy and momentum conservation  read $\omega= E-E'$ and $p_z'=-\omega \cos\theta$. Using (\ref{dispersion}) we obtain the radiated photon energies 
\begin{align} \label{eq:PhotonEnergyRotation}
\omega_0 =\frac{E-m'\Omega}{\sin^2\theta}
    \left\{ 1 - \left[ 1 - \frac{\mathcal{B} \sin^2\theta}{(E-m'\Omega)^2} \right]^{1/2} \right\}\,,
\end{align}
where we defined
\begin{align}\label{d11}
\mathcal{B} =& 2 (n-n') |qB| - \Omega^2 (m-m')^2 + 2 (E-m'\Omega) \Omega(m-m').
\end{align}

Unlike the unbounded case, in a finite cylinder we cannot analytically sum over the fermion polarizations $\zeta$, $\zeta'$ because the dependence on $\zeta'$ is not only contained in the coefficients $K_s$, but also in the principal quantum number $\lambda'$ and other quantities. Furthermore, since the principal quantum number $\lambda_i(\zeta,\,m)$ can be quite different for opposite polarizations, it is often misleading to compute the total radiation as an average of initial polarizations $\zeta=+$ and $\zeta=-$. We will compute instead the total radiation at fixed initial polarization.
The total intensity (at a given polarization $\zeta$) reads
\begin{equation}
\label{eq:exactformulaCylindricalBounded}
W= \sum_{h,\zeta'} W_{i,m,p_z=0,\zeta}^{h} =
 \frac{3}{2}\frac{W_\mathrm{Cl}}{E^2 (q B)^2}\int_0^\pi \D\theta \sum_{\zeta',\,m',\,i'}
\frac{\omega_0^2\sin\theta}{1+\frac{\omega_0\cos^2\theta}{E'-m'\Omega}} \Gamma^0_{i,m,\zeta}(i',m',\zeta',\theta),
\end{equation}
where
\begin{align}
\Gamma^0_{i,m,\zeta}(i',m',\zeta',\theta) = &
    \sum_h \mean{\,\vec{j}\cdot\vec{\Phi}\,} \nonumber\\
=& 2 \left[ K_1^2 \mathcal{J}_1^2
    + K_2^2 \mathcal{J}_2^2
    -2 K_1 K_2 \mathcal{J}_1\mathcal{J}_2\right. \nonumber\\
&+\cos^2\theta \left(K_1\mathcal{J}_1+ K_2 \mathcal{J}_2\right)^2
+\sin^2\theta \left(K_3\mathcal{J}_3 - K_4 \mathcal{J}_4\right)^2 \nonumber\\
&\left.+ \sin (2 \theta ) (K_1\mathcal{J}_1 + K_2\mathcal{J}_2) (K_3\mathcal{J}_3 - K_4\mathcal{J}_4)  \right],
\end{align}
and $W_\mathrm{Cl}$ is the classical intensity in the ultrarelativistic limit
\begin{align}\label{d35}
W_\mathrm{Cl}=\frac{q^2}{4\pi}\frac{2 (qB)^2E^2}{3}\,.
\end{align}
We use \eq{d35} to normalize  the radiation intensity presented in this work. Thus we consider the dimensionless quantity
\begin{equation}
\label{eq:WFinal}
\frac{W}{W_\mathrm{Cl}} = \sum_{h,\,\zeta'} \frac{W_{i,m,p_z=0,\zeta}^{h}}{W_\mathrm{cl}}
=
 \frac{3}{2}\frac{1}{E^2 (q B)^2}\int_0^\pi \D\theta \sum_{\zeta',\,m',\,i'}
\frac{\omega_0^2\sin\theta}{1+\frac{\omega_0\cos^2\theta}{E'-m'\Omega}} \Gamma^0_{i,m,\zeta}(i',m',\zeta',\theta),
\end{equation}
with the energy $E$ being the energy of the initial fermion
\begin{equation}
\label{eq:InitialE}
E=\sqrt{2\lambda_i(m,\,\zeta)|qB|+M^2}+m\Omega.    
\end{equation}
We will also compute the radiation intensity for a given photon's total angular momentum $l=m-m'$
\begin{equation}
\label{eq:Wlspectrum}
\frac{W(l)}{W_\mathrm{Cl}} = \sum_{h,\,\zeta'} \frac{W_{i,m,p_z=0,\zeta}^{h}(l)}{W_\mathrm{Cl}}
=
 \frac{3}{2}\frac{1}{E^2 (q B)^2}\int_0^\pi \D\theta \sum_{\zeta',\,i'}
\frac{\omega_0^2\sin\theta}{1+\frac{\omega_0\cos^2\theta}{E'-m'\Omega}} \Gamma^0_{i,m,\zeta}(i',m',\zeta',\theta)\,.
\end{equation}

Since the analytical expression of the roots $\lambda_i(m,\,\zeta)$ are not known, the summation on $\zeta'$, $m'$ and $i$ must be done numerically, as well as the integration over $\theta$. Moreover, in a finite cylinder the analytical results for the radial integrals in Eq.~(\ref{eq:RadialIntzCylinder}) are not known and must be evaluated numerically as well.

We will also compute the radiation intensity in the finite volume at vanishing magnetic field $B=0$ (``just rotation"). The corresponding expressions for the intensity are easily obtained by replacing the radial functions with the ones in Eq.~(\ref{eq:RadialFJR}) and with $i$ labeling the roots $p_{T,i}$ of the MIT BC equation (\ref{eq:MITConstraintJR}). 

%*********************************************************************************************************
\section{Numerical results}
\label{sec:Results}
%*********************************************************************************************************
In this section we present the procedure and the results of the numerical calculation of the radiation intensity (\ref{eq:WFinal}) and its angular momentum spectrum (\ref{eq:Wlspectrum}) for $\bar\sigma=$sgn$(qB)=-1$, e.g. an electron $q=-1$ in the magnetic field pointing in the $z$-direction $B >0$. The radiation intensity for a positive charge can be easily obtained by simply inverting the sense of rotation. The calculation is performed for two values of magnetic field $qB=-0.2 M^2$ and $qB=0$ and for a wide range of parameters: initial angular momentum $m$, angular velocity $\Omega$, and initial energy $E$ of Eq. (\ref{eq:InitialE}). The parameter that controls the influence of the boundary condition on the intensity is $\rho_R=|qB|/(2\Omega^2)$. We observed that at $\rho_R=100$ the system is in an intermediate stage where the boundary effects start to be significant. In addition, the calculation of radiation intensity can also be done in the unbounded case without breaking causality as long as the energy is low and the initial angular momentum is not large. In this case we compare the results obtained with unbounded solution and MIT BC.

In principle the summations in Eqs. (\ref{eq:WFinal}) and (\ref{eq:Wlspectrum}) must be done for each quantum number of the final fermion such that its energy is lower than the initial one. This is not a huge number of states when the rotation is negative, see Fig.~\ref{fig:energyLevelsM}, but the total number of final states can be very large for positive rotation, especially for small angular velocity, see Fig.~\ref{fig:energyLevelsP}. However, since the emission of photons with a large angular momentum is generally disfavored by the integrals (\ref{eq:RadialIntzCylinder}) inside the matrix element (\ref{AverPHI3Cylinder}), we do not need to include final states with an angular momentum $m'$ that is far from the initial angular momentum $m$ to obtain a result within a certain precision. Taking advantage of this feature the calculation has been performed as described in the following paragraphs. 

Having fixed the magnetic intensity $qB$, the angular velocity $\Omega$ and initial state with energy $E$, we solved the MIT BC (\ref{eq:MITConstraint}) and we computed the energy levels of the final fermions up to the energy $E$ with both polarizations $\zeta'=\pm$ and with an angular momentum $m'$ from $-|\bar{m}'|$ to $|\bar{m}'|$. Since the largest initial angular momentum that we considered is $m=\pm20.5$, we choose $|\bar{m}'|=100.5$ for $\rho_R=100$ and $|\bar{m}'|=50.5$ for smaller values of $\rho_R$. We then counted the total number of final states $N_f$ with $-|\bar{m}'|\leq m' \leq |\bar{m}'|$ and such that $E'\leq E$ and we computed the radiation intensity $W_f(\theta=\pi/2)$ for each of those single final states at $\theta=\pi/2$ where the intensity is at or close to its peak and we found the maximal state $\bar{f}$ giving the maximum of radiation intensity $\overline{W}_{\bar f}=$Max$\{W_f(\theta=\pi/2)\}$.
Given that we want to perform the calculation up to a relative precision $\epsilon$, among all these final states we only included the states such that they had a relevant contribution up to the precision $\epsilon$. We considered a final state $f$ a relevant state if it satisfies the following condition
\begin{equation}
    \frac{W_f(\theta=\pi/2)}{\overline{W}_{\bar f}} N_f > \epsilon.
\end{equation}
In other words, we are comparing the contribution from the state $f$ and the maximal contribution $\bar{f}$. A state $f$ is considered not relevant when the ratio between its contribution multiplied by the number of all considered states and the contribution from $\bar{f}$ is smaller then our goal precision. In this way since $f$ is a correction below the precision for $\bar{f}$,  it is also a correction below the precision for the total intensity. If the values of $m'$ between the relevant states did not reached the extreme values $\pm|\bar{m}'|$, the set of relevant states were considered enough to reach the precision $\epsilon$ and we proceeded to compute the summations within those states and the integration over $\theta$, otherwise we decreased the precision and we repeated the procedure.

For $\rho_R<100$ we were always able to reach a precision of $\epsilon=0.01$ and the error bars in the plots are smaller than the marker size used to represent the data point. For $\rho_R=100$ we were only able to reach the precision $\epsilon=0.01$ when the initial energy was lower than $E\sim 2 M$, for higher energies we decreased the precision when possible (and we show the error bars in the plots), however past a certain value the limit $|\bar{m}'|=100.5$ was not enough to obtain a meaningful result.

\subsection{Energy dependence}
\label{subsec:Energy}
%%%%%%%%%%%%%%%%%
\begin{figure}[th]
    \centering
    \includegraphics[width=.45\textwidth]{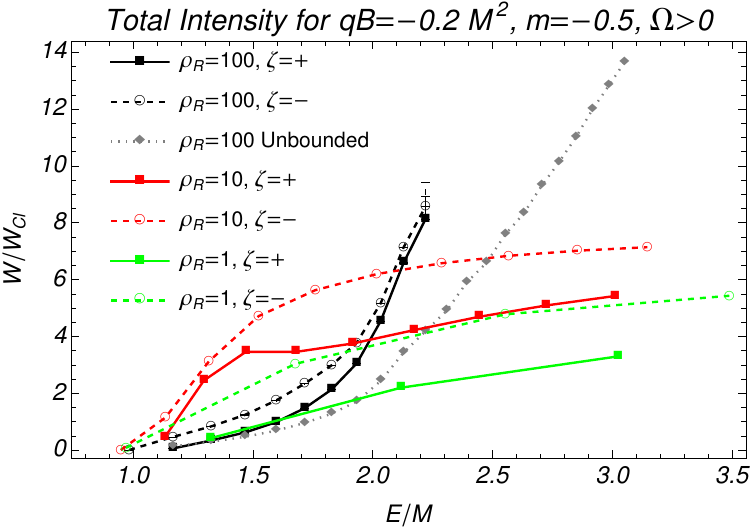}
    \includegraphics[width=.45\textwidth]{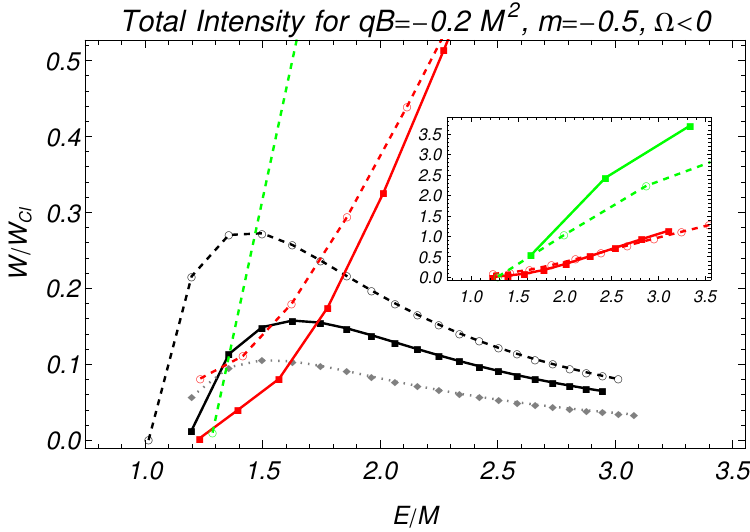}
    \caption{The total intensity of the synchrotron radiation  (\ref{eq:exactformulaCylindricalBounded}) in units of the classical intensity (\ref{d35}) as a function of the initial energy $E$ at $qB=-0.2 M^2$ and initial angular momentum $m=-0.5$ for various values of $\rho_R$, that is for different $\Omega$s. The gray dotted line with the diamond marker is averaged over the initial polarizations and is computed for $\rho_R=100$ with unbounded solution. All the other lines are obtained with MIT BC in a finite cylinder. Left: $\Omega>0$. Right:  $\Omega<0$.}
    \label{fig:EnergydepMITBC}
\end{figure}
%%%%%%%%%%%%%%%
%%%%%%%%%%%%%%%%%
\begin{figure}[th]
    \centering
    \includegraphics[width=.45\textwidth]{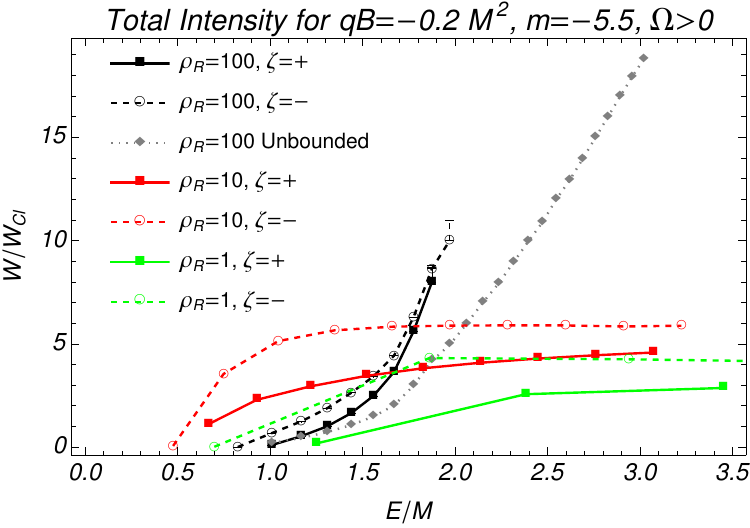}
    \includegraphics[width=.45\textwidth]{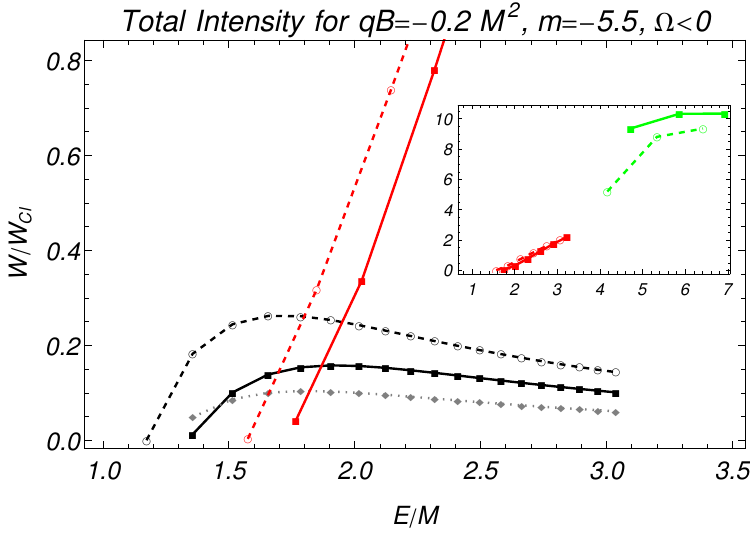}\\
    \includegraphics[width=.45\textwidth]{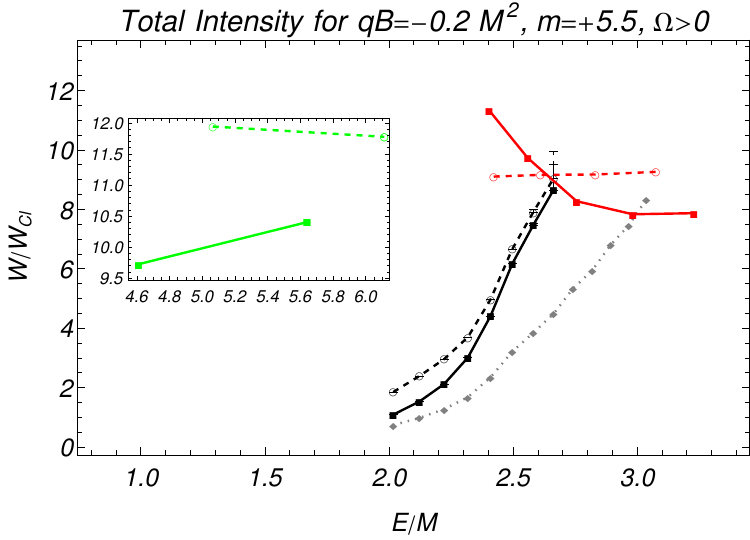}
    \includegraphics[width=.45\textwidth]{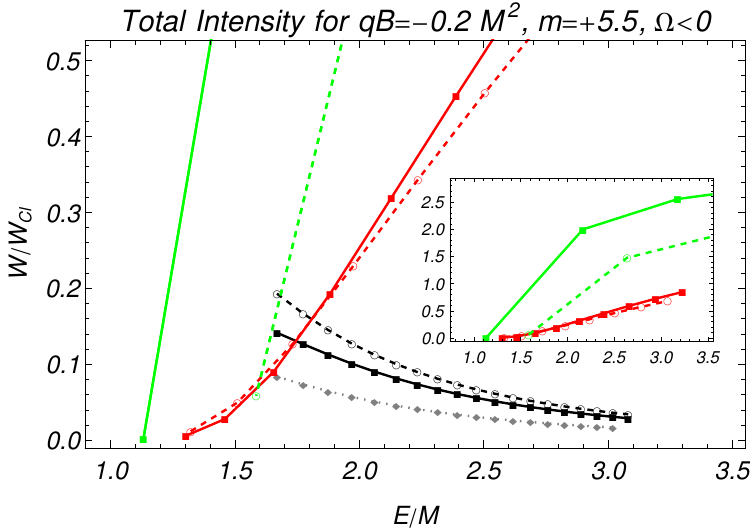}
    \caption{As in Fig.~\ref{fig:EnergydepMITBC}, the total intensity of the synchrotron radiation (\ref{eq:exactformulaCylindricalBounded}) in units of the classical intensity (\ref{d35}) as a function of the initial energy $E$ at $qB=-0.2 M^2$, at $m=-5.5$ (top), and $m=+5.5$ (bottom). Left: $\Omega>0$. Right:  $\Omega<0$.}
    \label{fig:EnergydepMITBC2}
\end{figure}
%%%%%%%%%%%%%%%%%
%%%%%%%%%%%%%%%%%
\begin{figure}[th]
    \centering
    \includegraphics[width=.45\textwidth]{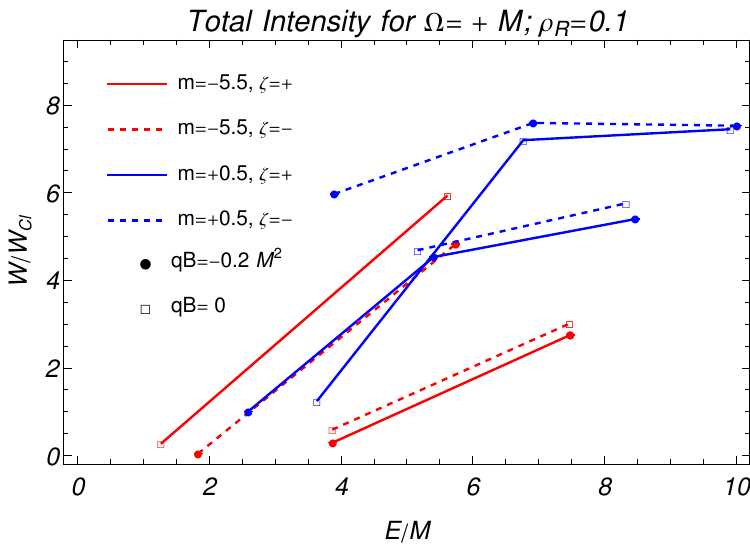}
    \includegraphics[width=.45\textwidth]{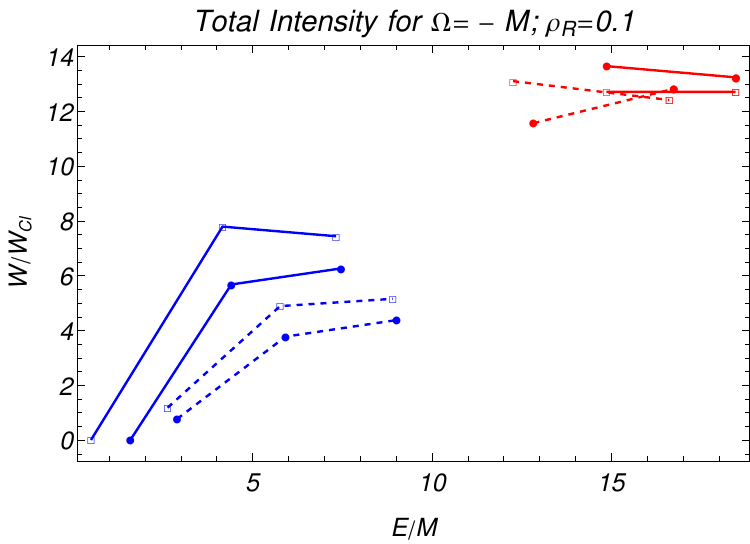}\\
    \includegraphics[width=.45\textwidth]{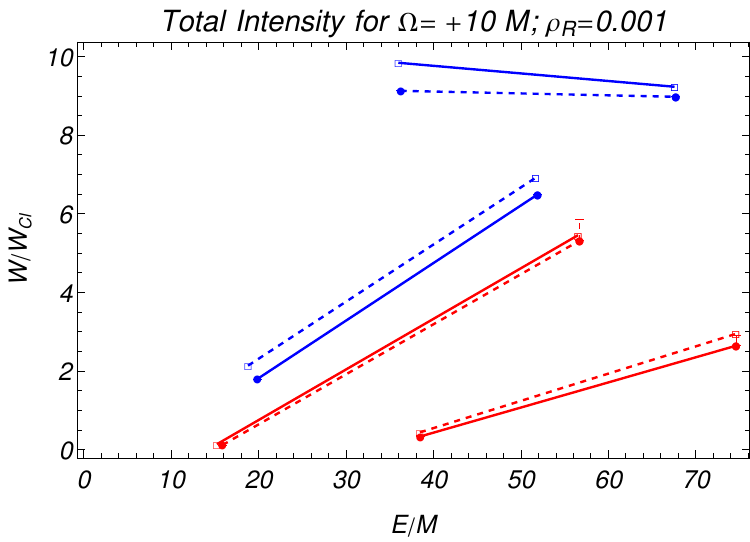}
    \includegraphics[width=.45\textwidth]{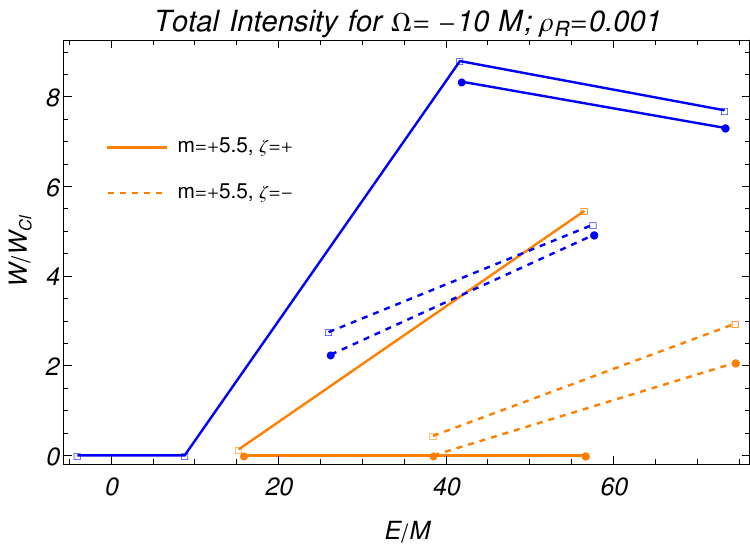}
    \caption{The total intensity of the synchrotron radiation  (\ref{eq:exactformulaCylindricalBounded}) in units of the classical intensity (\ref{d35}) as a function of the initial energy $E$ at $qB=-0.2 M^2$ (filled circles) and at $qB=0$ (empty squares) and for various $m$. Solid lines corresponds to $\zeta=+$, dashed ones to $\zeta=-$. Top: $|\Omega| = M,\, \rho_R=0.1$. Bottom: $|\Omega| =10 M,\,\rho_R=0.001$. Left: $\Omega>0$. Right: $\Omega<0$.}
    \label{fig:EnergydepMITBCFR}
\end{figure}
%%%%%%%%%%%%%%%%%

In \cite{Buzzegoli:2022dhw,Buzzegoli:2023vne} the total synchrotron radiation intensity (averaged over initial polarization) has been computed for slow rotation in the unbounded case. It was found that for negative $qB<0$, a positive rotation $\Omega >0$ results in an enhancement of radiation compared with the case without rotation, that is $W/W_{\rm Cl}\gtrsim 1$, while a negative rotation $\Omega <0$ results in a suppression of radiation, that is $W/W_{\rm Cl}\lesssim 1$. One of the questions addressed in this work is if this behavior is preserved for fast rotation.

The dependence of total radiation intensity $W$, obtained from Eq.~(\ref{eq:exactformulaCylindricalBounded}), on the initial energy is illustrated in Fig.~\ref{fig:EnergydepMITBC} for $m=-0.5$ and various values of positive and negative rotations, and for $m=\pm5.5$ in Fig.~\ref{fig:EnergydepMITBC2}. The case $m=+0.5$ is similar to Fig.~\ref{fig:EnergydepMITBC}. For positive rotation the effect still results in an enhancement. However, as we increase the angular velocity, the factor of enhancement is not increasing with angular velocity, it instead decreases past some value of $\Omega$. It also shows a complicated dependence on $m$ and $\rho_R$ which is difficult to infer from purely qualitative analysis without making the calculation, see for instance how for $m=+5.5$ and $\rho_R=10$ (red lines of the bottom left plot of Fig.~\ref{fig:EnergydepMITBC2}) the ratio $W/W_{\rm Cl}$ is flat or decreasing with energy, meaning that $W$ is scaling slower than $W\sim E^2$. For negative rotation, the intensity is suppressed until the energy and $\Omega$ are large enough. This is easily explained if we compare the synchrotron frequency $\omega_B=|qB|/E$, which gives the scale for synchrotron radiation, with the angular velocity $\Omega$. As we increase $E$ or $\Omega$ the synchrotron frequency becomes smaller than the angular velocity. Therefore, the dominant process is the radiation from the global rotation of the system and the synchrotron radiation from the magnetic field is a correction.

Notice that for different values of $\rho_R$, $\zeta$ and $m$ the minimum energy can be vastly different and that it can also be lower than the mass: $E<M$. This is an effect of the BC, as shown in the energy spectrum of Figs~\ref{fig:energyLevelsP} and~\ref{fig:energyLevelsM}. Figure~\ref{fig:EnergydepMITBC} and Fig.~\ref{fig:EnergydepMITBC2} for $\rho_R=100$ also show the direct comparison between the unbounded and the MIT BCs. We observe that the MIT BC have a strong dependence on the polarization and for both positive and negative rotation the intensity is higher. We also point out that for positive rotation, the results obtained with the unbounded solution at high energy are not physical. Indeed, in those cases the quantum numbers $n',\,a'$ may exceed the value $\rho_R=100$, resulting in fermions with an average radial radius $\mean{r^2}>\Omega^{-2}$ and hence breaking causality.

Figure~\ref{fig:EnergydepMITBCFR} shows the energy dependence of the intensity for very fast rotation, i.e. $|\Omega| = M$ ($\rho_R=0.1$) and $|\Omega| =10 M$ ($\rho_R=0.001$).
In this case we expect the effect of the magnetic field to be a small correction compared to rotation. Indeed, comparing with the radiation intensity of a rotating fermion in absence of magnetic field the results are quite close. In this case the energies are large because the volume is small. For positive rotation and negative $qB$ the intensity of $qB=0$ matches with those of finite $qB$ with opposite polarization, while for negative rotation with those with the same polarization. That is because in the latter case we flipped the direction of rotation but not that of the reference frame from which we define the polarization.

\subsection{Initial angular momentum dependence}
\label{subsec:Initialm}
%%%%%%%%%%%%%%%%%
\begin{figure}[th]
    \centering
    \includegraphics[width=.45\textwidth]{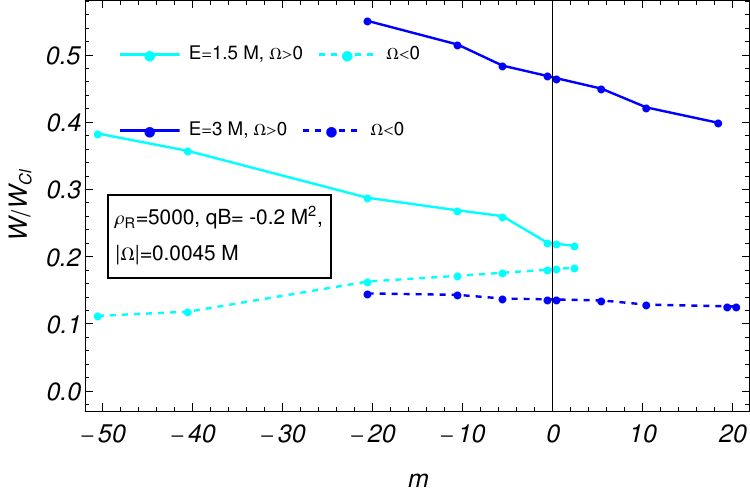}
    \includegraphics[width=.45\textwidth]{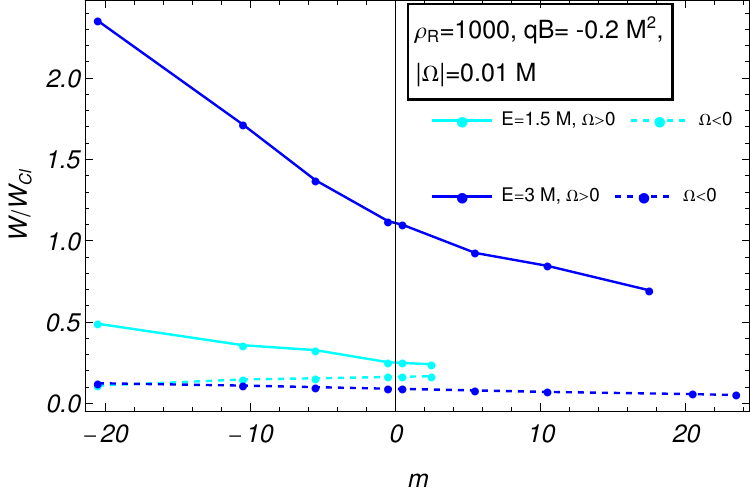}\\
    \caption{The total intensity of the synchrotron radiation (\ref{eq:exactformulaCylindricalBounded}) computed with unbounded solution in units of the classical intensity (\ref{d35}) as a function of the initial angular momentum $m$ at $qB=-0.2 M^2$ and small angular velocities. Cyan lines are for an initial energy $E/M\simeq 1.5$ and blue lines for $E/M\simeq 3$. Solid lines are for $\Omega>0$ and dashed lines are for $\Omega<0$. Left: $\rho_R=5000$. Right: $\rho_R=1000$. }
    \label{fig:AMdepUnb}
\end{figure}
%%%%%%%%%%%%%%%
%%%%%%%%%%%%%%%%%
\begin{figure}[th]
    \centering
    \includegraphics[width=.45\textwidth]{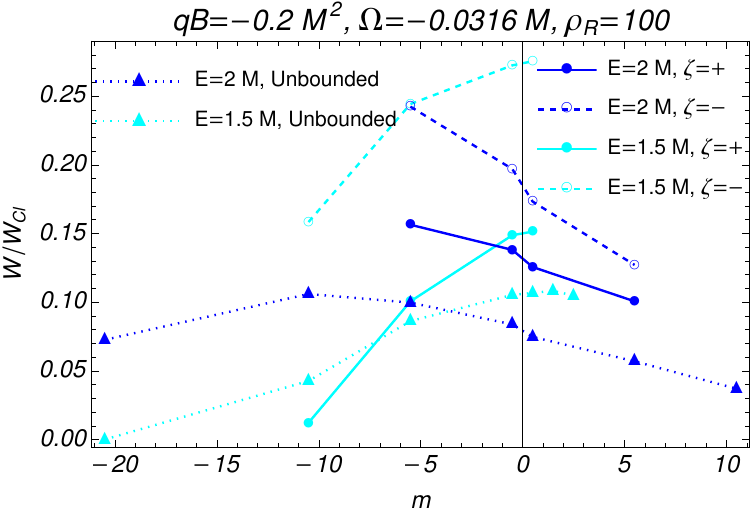}
    \includegraphics[width=.45\textwidth]{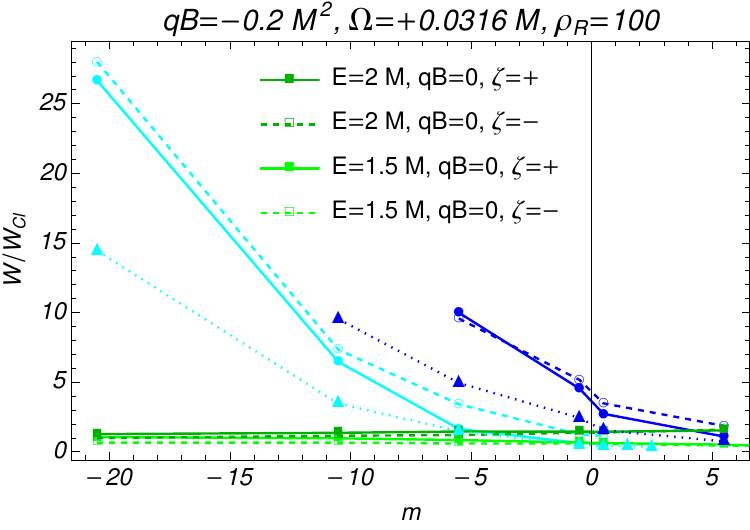}\\
    \includegraphics[width=.45\textwidth]{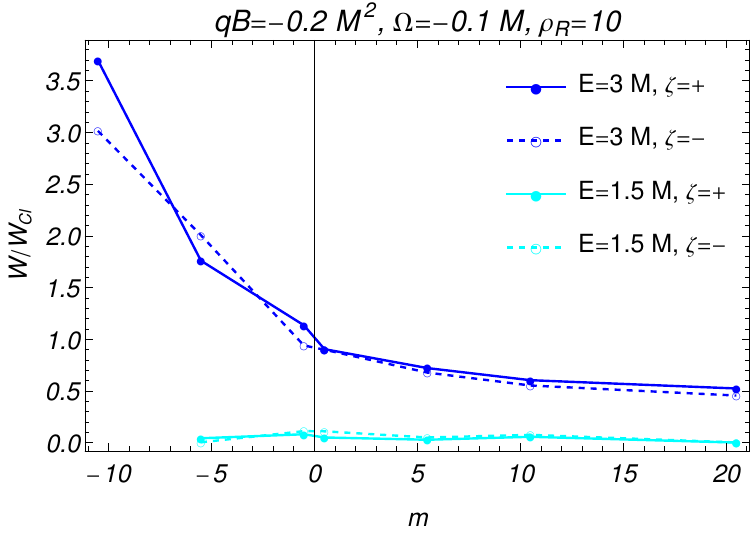}
    \includegraphics[width=.45\textwidth]{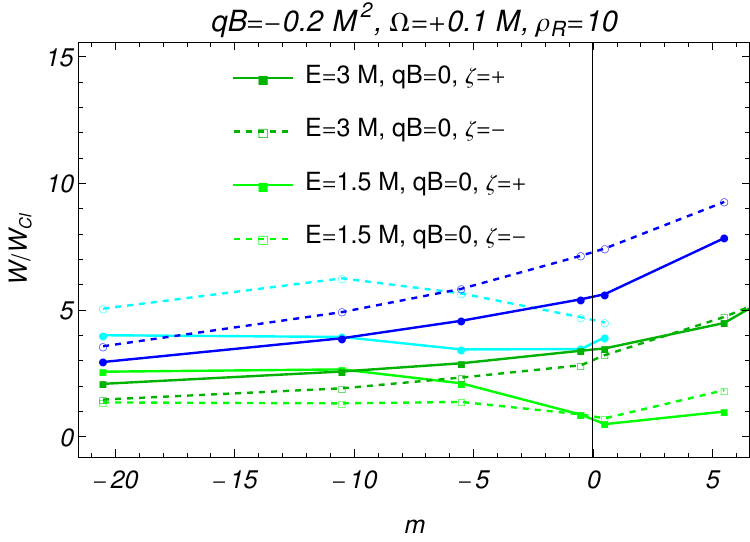}
    \caption{The total intensity of the synchrotron radiation (\ref{eq:exactformulaCylindricalBounded}) computed with unbounded (triangles) and MIT BC (circles and squares) in units of the classical intensity (\ref{d35}) as a function of the initial angular momentum $m$ at $qB=-0.2 M^2$ (circles) and at $qB=0$ (squares). Top: $\rho_R=100$. Bottom: $\rho_R=10$.}
    \label{fig:AMdepMITBC}
\end{figure}
%%%%%%%%%%%%%%%
%%%%%%%%%%%%%%%%%
\begin{figure}[th]
    \centering
    \includegraphics[width=.45\textwidth]{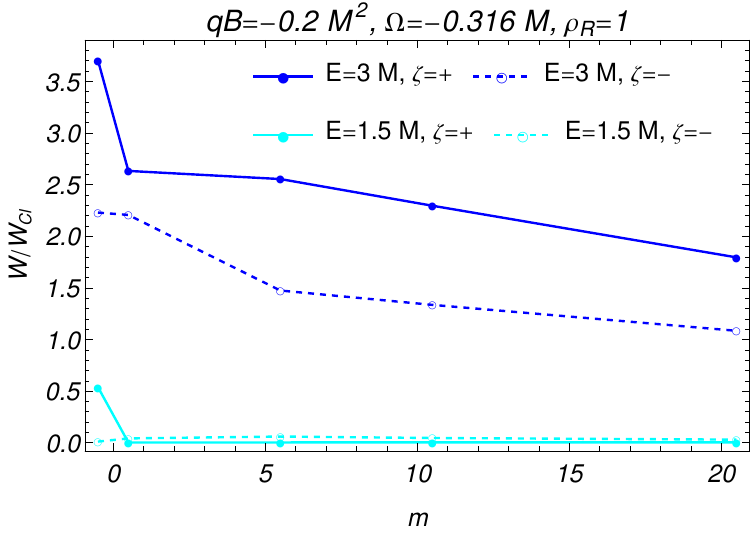}
    \includegraphics[width=.45\textwidth]{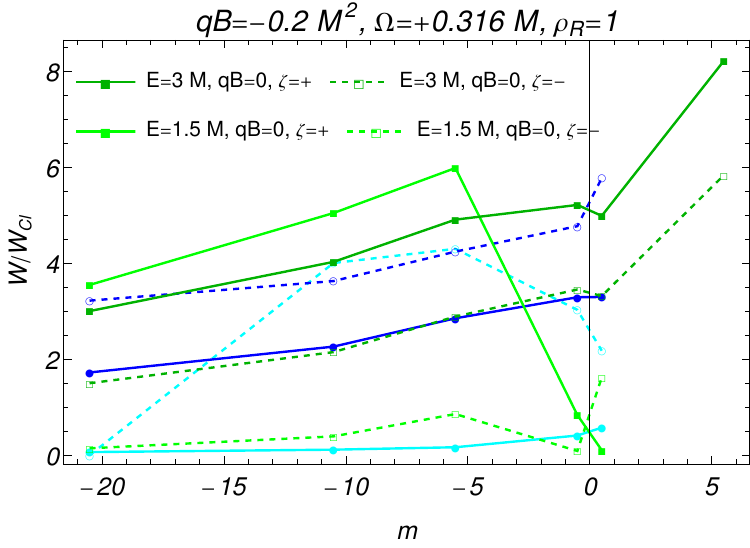}\\
    \includegraphics[width=.45\textwidth]{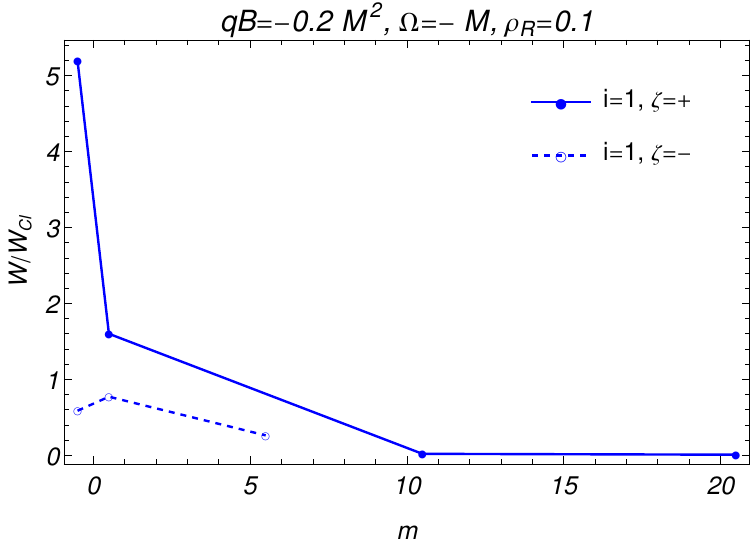}
    \includegraphics[width=.45\textwidth]{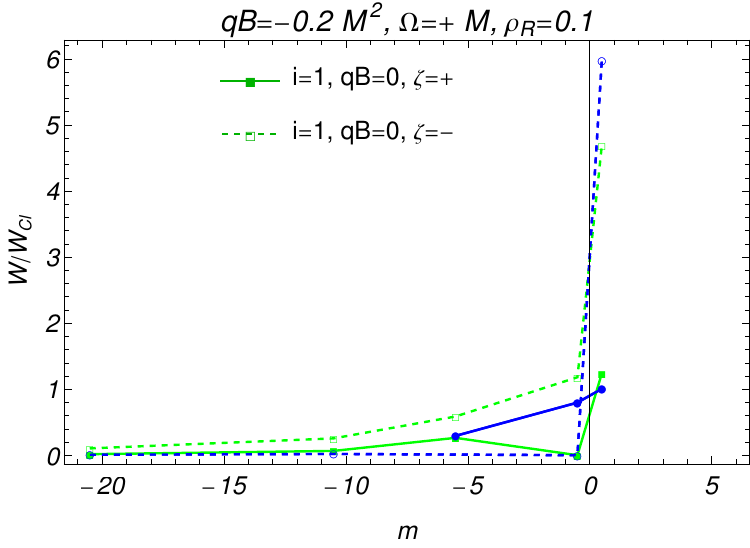}
    \caption{The total intensity of the synchrotron radiation (\ref{eq:exactformulaCylindricalBounded}) computed with MIT BC in units of the classical intensity (\ref{d35}) as a function of the initial angular momentum $m$ at $qB=-0.2 M^2$ (circles) and at $qB=0$ (squares). Top: $\rho_R=1$. Bottom: $\rho_R=0.1$, $i=1$ denotes the first energy level at given $\zeta$ and $m$. }
    \label{fig:AMdepMITBC2}
\end{figure}
%%%%%%%%%%%%%%%
From Figs.~\ref{fig:EnergydepMITBC}, \ref{fig:EnergydepMITBC2} and \ref{fig:EnergydepMITBCFR}, we found that the radiation intensity emitted by a single fermion can strongly depend on its angular momentum $m$. To study this dependence we are showing how the radiation intensity at fixed initial energy $E$ changes with $m$. We choose $E=1.5 M$ and $E= 3M$ for all values of $\rho_R$, except for $\rho_R=100$ for which we choose $E=1.5 M$ and $E= 2M$ because in this case larger energies require a steep increase in computational time. It must be noted that the energy $E$ also depends on $m$, $\zeta$ and $\rho_R$, and the initial fermion state cannot be chosen exactly at the energy we want. The energies of the data points in the lines of constant energy in Figs.~\ref{fig:AMdepUnb}, \ref{fig:AMdepMITBC} and \ref{fig:AMdepMITBC2} actually fluctuates around the reported values but they never deviates more than $20\%$. For the fast rotation case $\rho_R=0.1$ the energies change drastically at different $m$'s, therefore we considered the first energy level at given $m$ and $\zeta$.

The usual synchrotron radiation in absence of rotation does not depend on $m$. When rotation is introduced, the degeneracy in angular momentum is lost. However, for small angular velocities this dependence is mild. Figure~\ref{fig:AMdepUnb} shows the slow rotation case with unbounded solution. The largest impact on intensities is due to the sign of rotation which determines if the radiation will be enhanced or suppressed. A significant dependence on $m$ starts to appear for positive rotation and large energies. Notice that the unbounded solution requires that the radial quantum number $a=n-m-1/2$ must be non-negative. For this reason not all values of $m$ are allowed at a given energy.

The case $\rho_R=100$ is illustrated in Fig.~\ref{fig:AMdepMITBC} and the results with unbounded and MIT BCs are compared. The results for $\rho_R=100$ with MIT BC, $E= 2M$ and $|m|>5.5$ require much more computation time than the other points and have not been obtained. For negative rotation we see a huge difference between the infinite and the finite volume, while for positive rotation they have the same qualitative behavior. As noted for the energy dependence, the intensity computed with MIT BC is larger. For positive rotation we are also comparing with the radiation at $qB=0$. The  combined effect of rotation and magnetic field is much higher than the sum of the two alone. For $\rho_R=10$ (on the bottom of Fig.~\ref{fig:AMdepMITBC}) the radiation intensities from finite and vanishing magnetic field approach each other. We note that for positive rotation and $\rho_R=100$ the intensity is larger for large and negative $m$, while the opposite trend is found for $\rho_R=10$. The initial angular momentum dependence for higher angular velocities is shown in Fig.~\ref{fig:AMdepMITBC2}.

This discussion has an interesting consequence for photon production by the rotating quark-gluon plasma. Even though the states with same energy $E$ presented in the Figs.~\ref{fig:AMdepUnb}, \ref{fig:AMdepMITBC} and \ref{fig:AMdepMITBC2} populate the plasma with the same density, they contribute differently to the electromagnetic radiation.

\subsection{Angular velocity dependence}
\label{subsec:Rotation}
%%%%%%%%%%%%%%%%%
\begin{figure}[th]
    \centering
    \includegraphics[width=.45\textwidth]{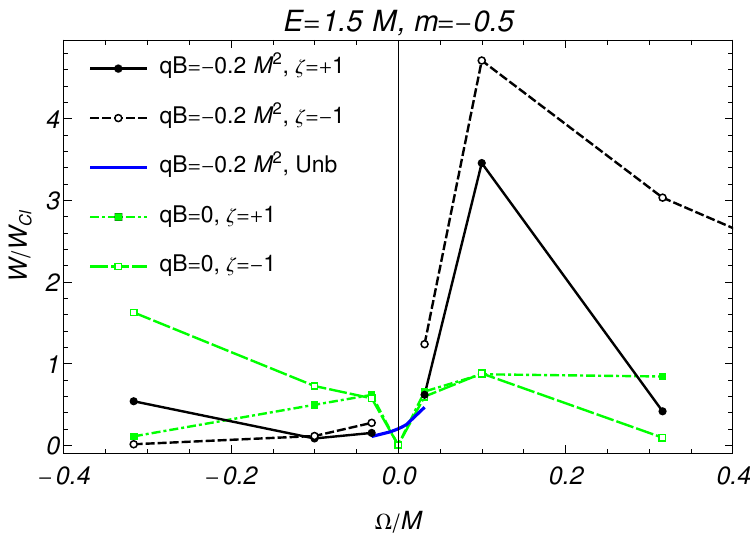}
    \includegraphics[width=.45\textwidth]{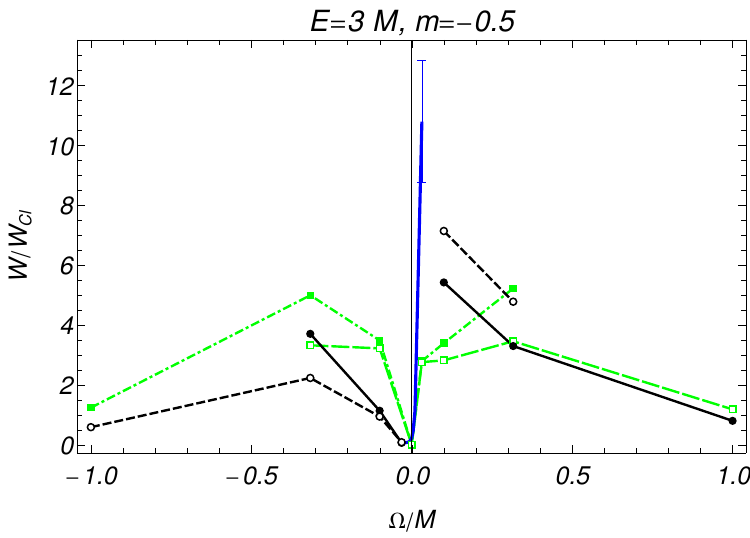}\\
    \includegraphics[width=.45\textwidth]{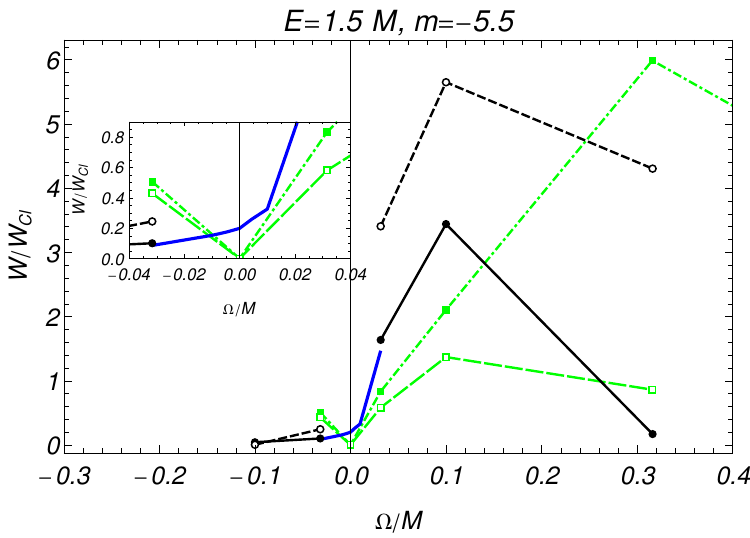}
    \includegraphics[width=.45\textwidth]{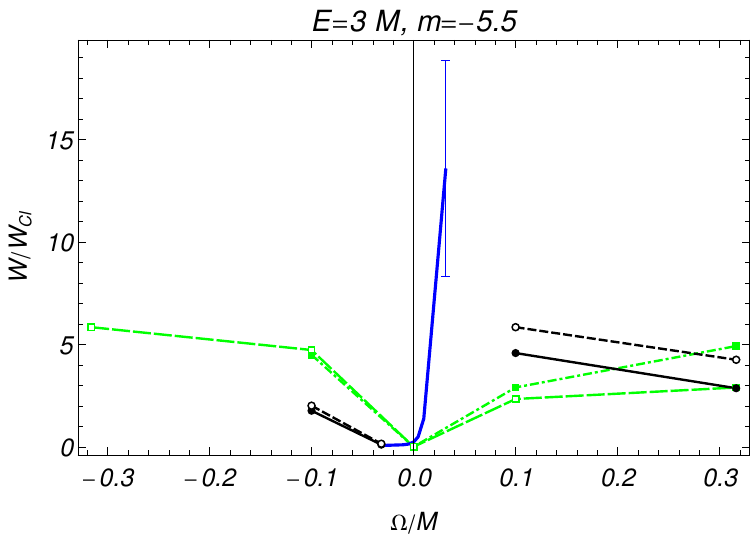}
    \caption{The total intensity of the synchrotron radiation (\ref{eq:exactformulaCylindricalBounded}) in units of the classical intensity (\ref{d35}) as a function of the angular velocity $\Omega$ for fixed values of $m$; Top: $m=-0.5$. Bottom: $m=-5.5$. Black lines are obtained with MIT BC at $qB=-0.2 M^2$, green lines at $qB=0$. The blue line is the intensity averaged over polarization obtained with unbounded solution valid at small $\Omega$. The error bar indicates the result imposing the causality constraint (lower value) and without imposing it (higher value).}
    \label{fig:WVsRot}
\end{figure}
%%%%%%%%%%%%%%%%%

Let us now examine in more detail Figs.~\ref{fig:EnergydepMITBC}, \ref{fig:EnergydepMITBC2} and \ref{fig:EnergydepMITBCFR} where  we found that increasing the angular velocity for positive rotation reduced the enhancement factor and for negative rotation reduced the suppression factor. As done in the analysis of the initial angular momentum, we fix the energy of the initial state and its angular momentum and we study the radiation intensity for different values of the angular velocity $\Omega$. The results are shown in Fig.~\ref{fig:WVsRot}. For small values of $\Omega$, i.e.\ for $\rho_R\geq 100$, we used unbounded solution, since for large $\rho_R$ the effect of the boundary is minor. In this case the total intensity is given averaging the positive and negative polarizations and is shown in blue. In the case of positive rotation and $E=3 M$ some of the final states did not satisfy the causality condition $n'+a'\leq\rho_R=100$. In this case we report two results, one obtained including only the casual states $n'+a'\leq\rho_R$, whose value is located at the bottom of the error bar, and one including all states which is reported at the top of the error bar. For $\rho_R=100$ we also compare with the result obtained with MIT BC (except for $\Omega>0$ and $E= 3 M$ as explained earlier). The black lines are the values for MIT BC and are given separately for $\zeta=+$ and for $\zeta=-$. The green lines show the radiation intensity for a rotating system in the absence of magnetic field obtained with MIT BC.

Figure~\ref{fig:WVsRot} confirms the behavior described above and we found the same qualitative behavior for other values of $m$. For positive rotation we observe a rapid enhancement of radiation for small and moderate $\Omega$, followed by a sudden decrease in the enhancement factor for larger angular velocities. The decrease in the enhancement can be understood noticing that we are fixing the energy $E$ and that by increasing $\Omega$, the lowest energy level is raised due to the boundary effects. This implies that the initial fermion state has less states where to decay, resulting in a lower radiation intensity.

For negative rotation instead, at rapid rotation the radiation inverts the trend and starts increasing instead of decreasing because at large $\Omega$ the effect of rotation is dominant compared to the magnetic field and it is approaching the radiation intensity at $qB=0$. Indeed, for both positive and negative rotation the results at finite magnetic field and at $qB=0$ starts to merge at large rotation. The dominance of rotation is also the reason why at large negative $\Omega$ the intensity from $\zeta=-$ fermions starts to be lower than the one from $\zeta=+$, see the left plots of Fig.~\ref{fig:WVsRot}. At $qB=0$ and for $\Omega=0$ there is no radiation because the fermion is not accelerating, while at finite $qB$ the fermion is emitting the well known synchrotron radiation.

\subsection{Photon's angular momentum spectrum}
\label{subsec:lSpectrum}
%%%%%%%%%%%%%%%%%
\begin{figure}[th]
    \centering
    \includegraphics[width=.45\textwidth]{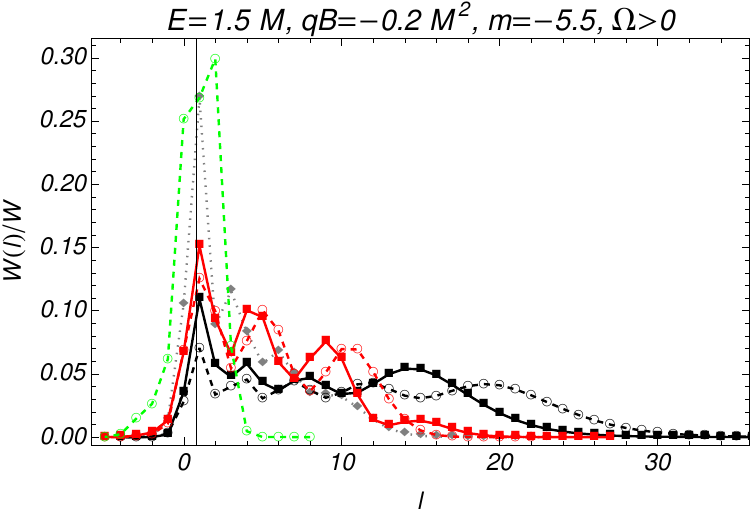}
    \includegraphics[width=.45\textwidth]{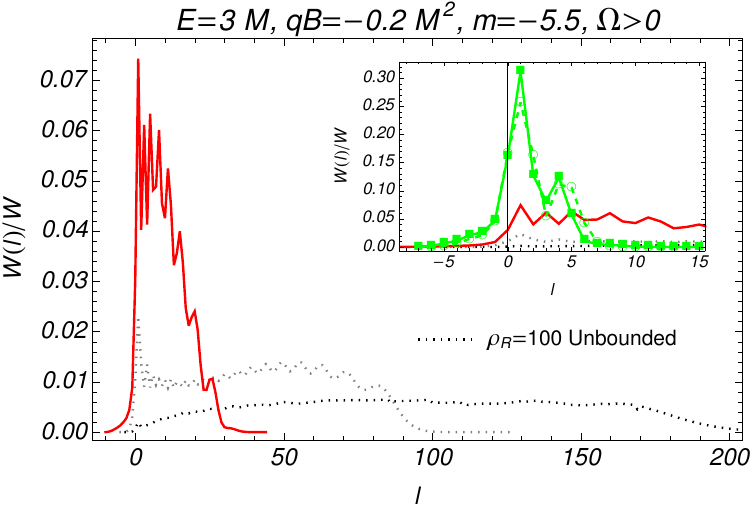}\\
    \includegraphics[width=.45\textwidth]{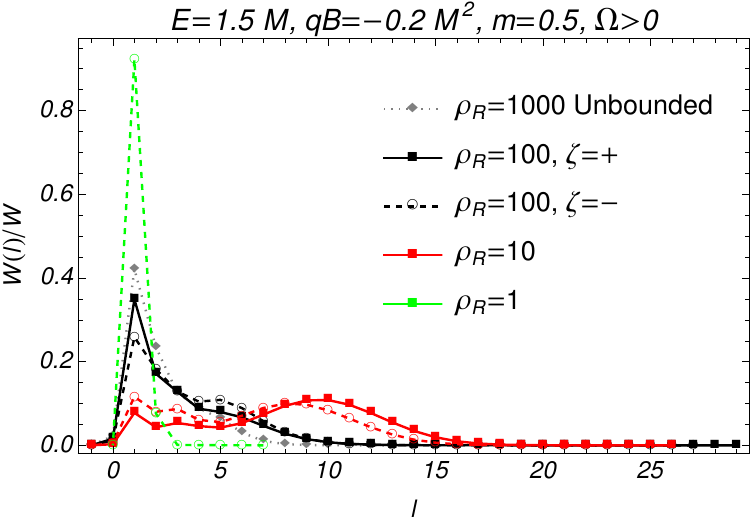}
    \includegraphics[width=.45\textwidth]{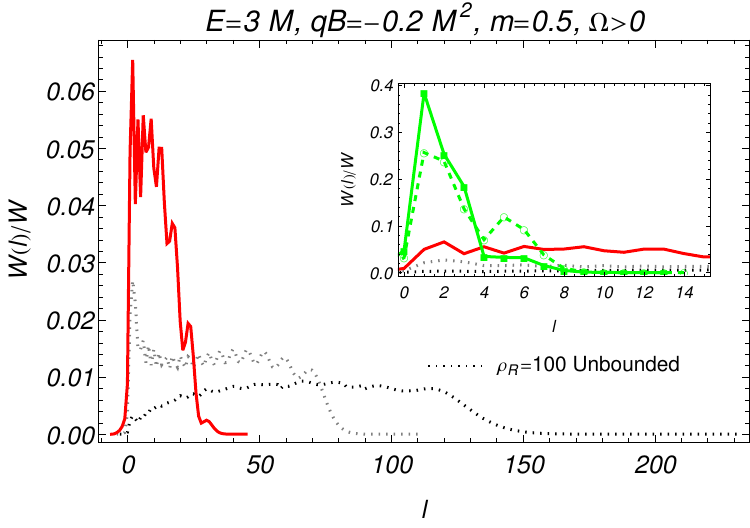}
    \caption{The total radiation spectrum in the photon's angular momentum (\ref{eq:Wlspectrum}) in  units of the classical intensity (\ref{d35}) at $qB=-0.2 M^2$ for positive rotation $\Omega>0$. In the horizontal axis $l=m-m'$ is the angular momentum of the photon and in the vertical axis the corresponding radiation intensity. Dotted lines denotes the unbounded solution, solid lines MIT BC with $\zeta=+$ and dashed line MIT BC with $\zeta=-$. Different colors correspond to different angular velocities $\Omega$. Top: $m=-5.5$. Bottom: $m=+0.5$. Left: $E= 1.5 M$. Right: $E= 3 M$.}
    \label{fig:PAMSpectrumP}
\end{figure}
%%%%%%%%%%%%%%%%%
%%%%%%%%%%%%%%%%%
\begin{figure}[th]
    \centering
    \includegraphics[width=.45\textwidth]{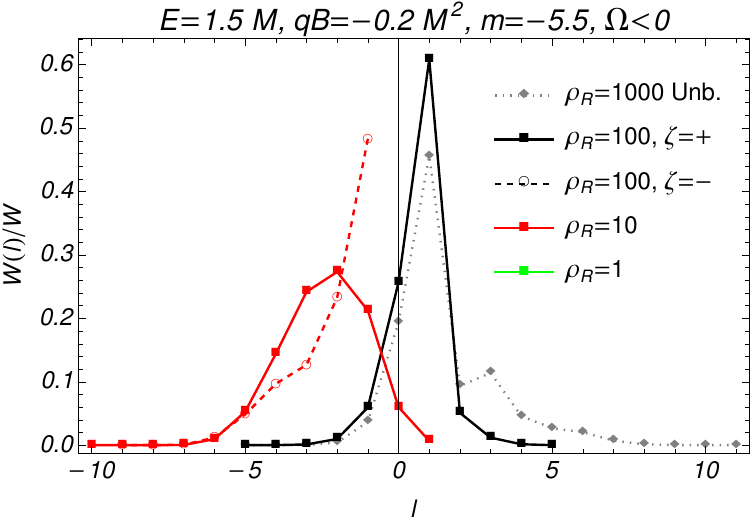}
    \includegraphics[width=.45\textwidth]{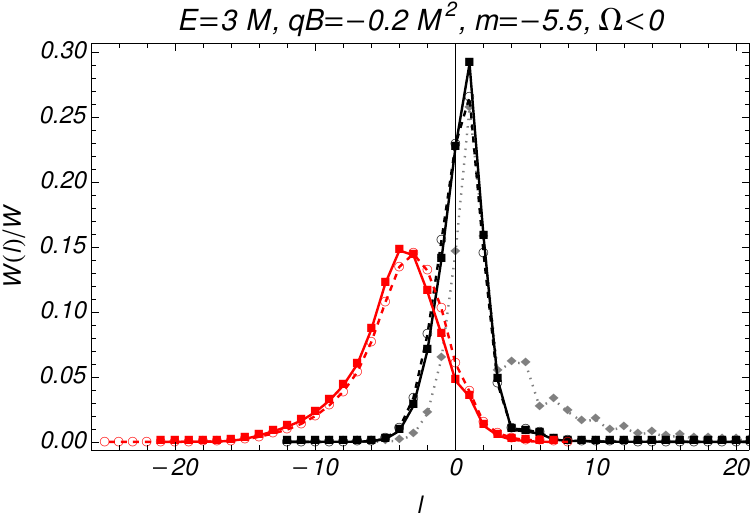}\\
    \includegraphics[width=.45\textwidth]{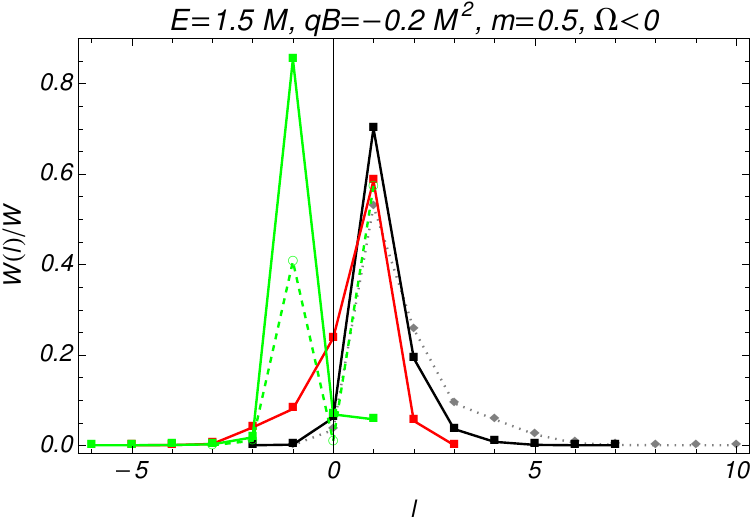}
    \includegraphics[width=.45\textwidth]{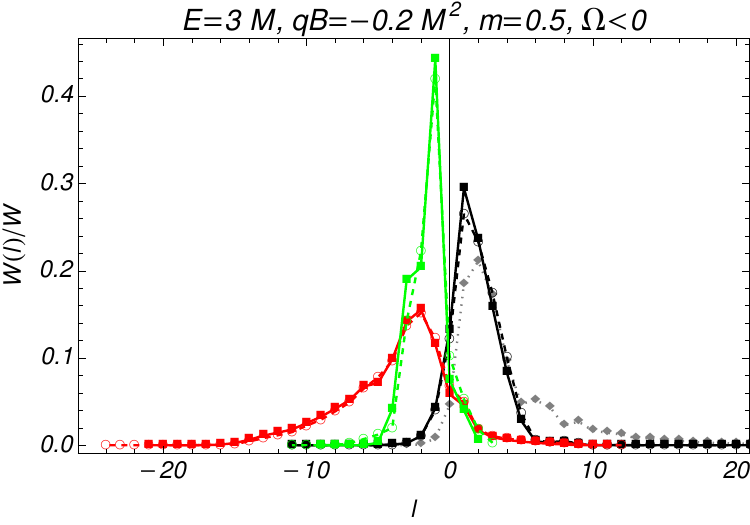}
    \caption{The total radiation spectrum in the photon's angular momentum (\ref{eq:Wlspectrum}) in  units of the classical intensity (\ref{d35}) at $qB=-0.2 M^2$ as in Fig.~\ref{fig:PAMSpectrumP} but for negative rotation $\Omega<0$. Top: $m=-5.5$. Bottom: $m=+0.5$. Left: $E= 1.5 M$. Right: $E= 3 M$.}
    \label{fig:PAMSpectrumM}
\end{figure}
%%%%%%%%%%%%%%%%%
%%%%%%%%%%%%%%%%%
\begin{figure}[th]
    \centering
    \includegraphics[width=.45\textwidth]{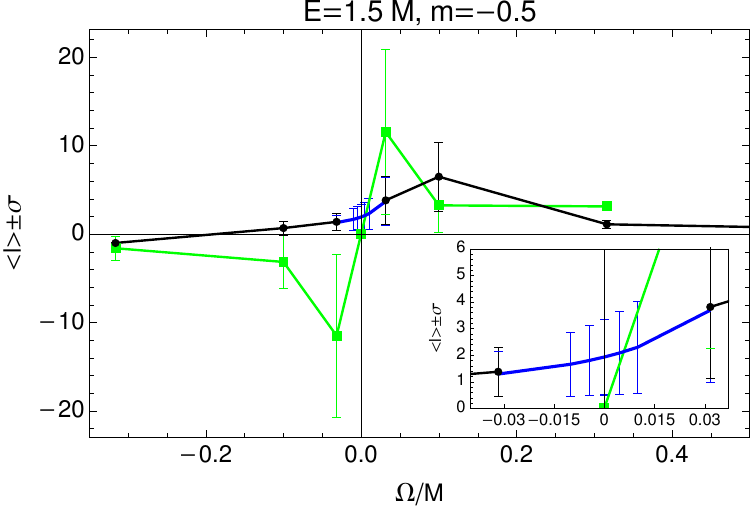}
    \includegraphics[width=.45\textwidth]{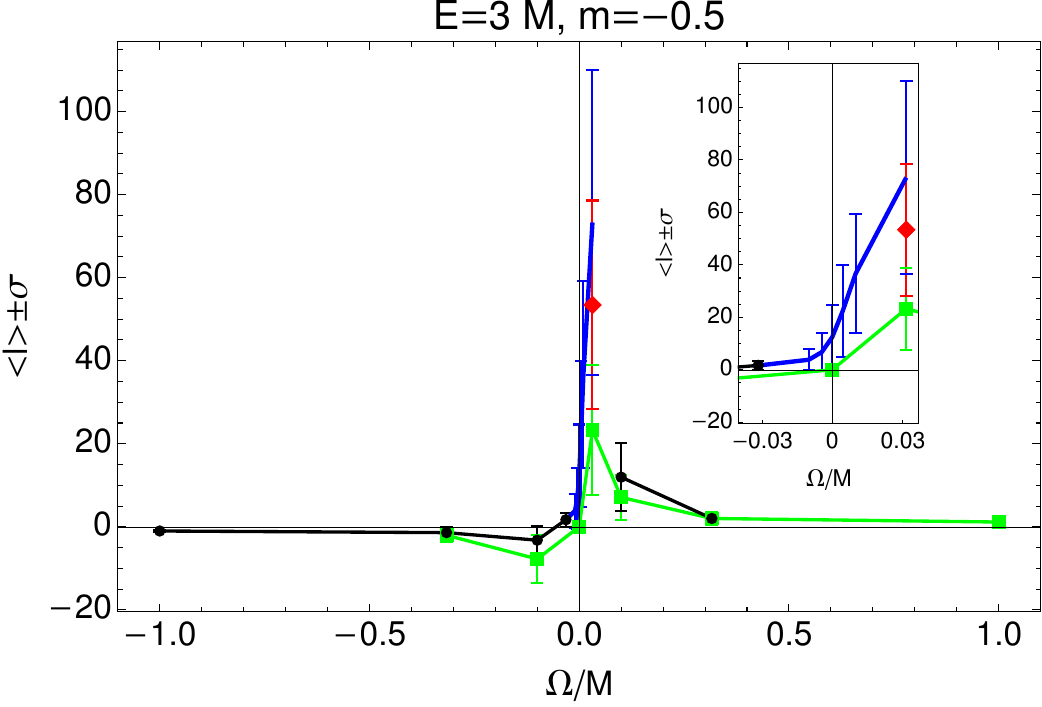}\\
    \includegraphics[width=.45\textwidth]{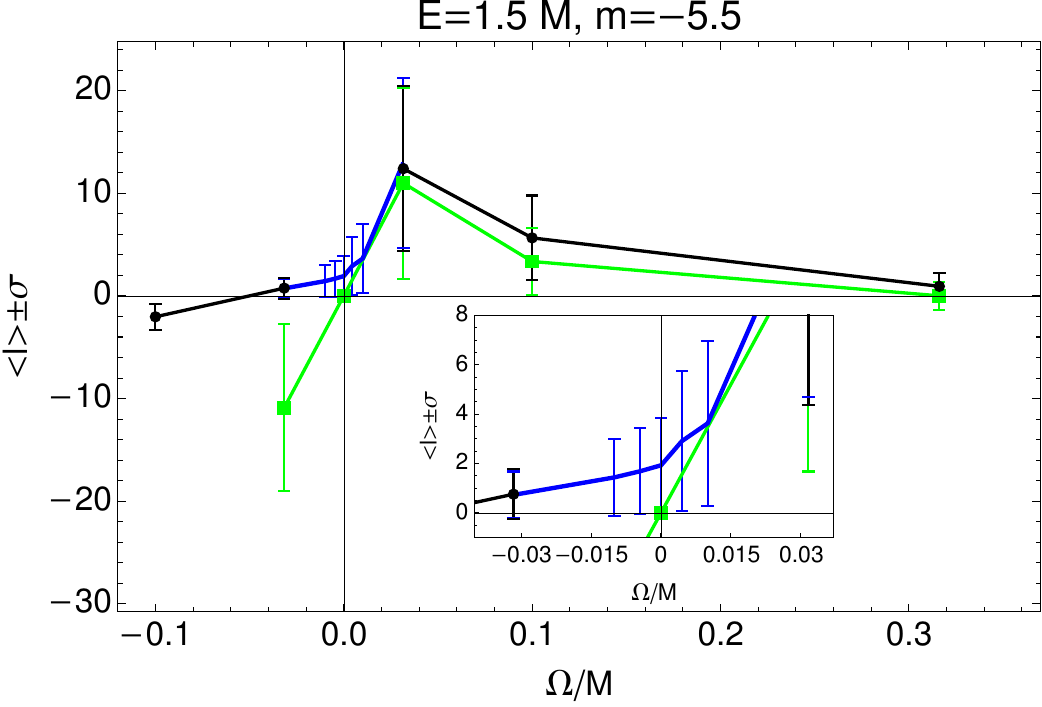}
    \includegraphics[width=.45\textwidth]{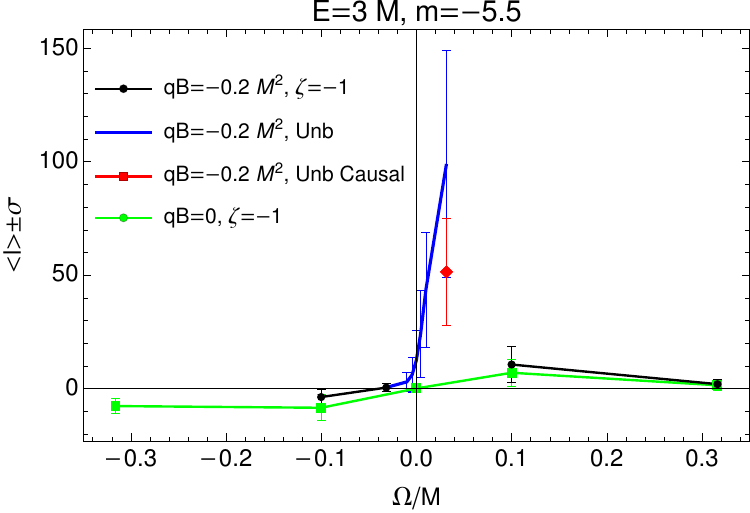}
    \caption{The average photon angular momentum $\mean{l}$ and the standard deviation $\sigma$, given as an error bar, of the radiation intensity spectrum (\ref{eq:Wlspectrum}) at $qB=-0.2 M^2$. The blue line is the unbounded case, the red point is obtained in the unbounded case including only casual states. The black line is with MIT BC from $\zeta=-$ and the green line is at $qB=0$ with MIT BC from $\zeta=-$. Top: $m=-0.5$. Bottom: $m=-5.5$. Left: $E= 1.5 M$. Right: $E= 3 M$.}
    \label{fig:PAMVsRot}
\end{figure}
%%%%%%%%%%%%%%%%%
Since the angular velocity is canonical conjugate to the angular momentum, we investigated how the angular momentum of the emitted radiation is distributed. This analysis  allows us to differentiate a radiation caused by a magnetic field from radiation from a rotating system and to find the direction of $qB$ and $\Omega$ by measuring the angular momentum of the photons. It also indicates the emission probability of photons with high angular momentum, which has attracted certain interest in the literature (see e.g.\ \cite{Ivanov:2022jzh} and references therein). 

The process of the emission of a photon conserves the angular momentum and the photon angular momentum is the difference between the angular momentum of the initial and final fermion $l=m-m'$. The radiation intensity from a fixed initial state to all possible states with the photon's angular momentum $l$ is given in Eq.~(\ref{eq:Wlspectrum}).
We show the spectrum for the initial states with $m=+0.5$ and $m=-5.5$ and with energies $E= 1.5 M$ and $E=3 M$ for positive rotation in Fig.~\ref{fig:PAMSpectrumP} and for negative rotation in Fig.~\ref{fig:PAMSpectrumM}. As a reference we also computed the spectrum for a large values of $\rho_R$ (small angular velocity $\Omega$), i.e. $\rho_R=1000$. In this case we used the unbounded solution because the effect of the boundary is not important. For $E= 3M$, $\Omega>0$ and $\rho_R=100$ we used the results from the unbounded solution instead of the MIT BC.

For positive rotation we observe an oscillatory behavior and the emission of photons with angular momentum in the opposite direction of $qB$, i.e. $l>0$. For larger initial energy at small $\Omega$ the radiation contains photons with a wide range of $l$ while for fast rotation the distribution is much narrower and centered at low positive $l$. Counter-intuitively, we found that the largest angular momenta are obtained for moderate angular velocities.
For negative rotation the spectrum is narrower and it is centered at negative angular momentum ($l<0$) only for fast rotation. We found similar qualitative behavior for other values of $m$. The preference for positive $l$ in magnetic field is simpler to be explained in the unbounded case. For $\bar\sigma=$sgn$(qB)=-1$, the fermionic angular momentum is $m=n-a-\tfrac{1}{2}$ with $n\geq 0$ and $a\geq 0$ and cannot be higher than $m=n-\tfrac{1}{2}$ but can have any lower value. Fixing the initial $m$, we generally have more final fermionic state with $m'<m$ than with $m'>m$ resulting in a bias for $l>0$. The opposite occurs for $\bar\sigma=+1$, implying that we can measure the radiation angular momentum to infer the direction of $qB$.

To illustrate how the spectrum changes at fixed $qB$ and varying $\Omega$, we computed the average photon angular momentum of the spectrum, defined as
\begin{equation}
\mean{l} = \sum_l \frac{l\, W(l)}{W} = \sum_{m'} \frac{(m-m') W(m')}{W},
\end{equation}
with $W(l)$ the intensity at a given final angular momentum (\ref{eq:Wlspectrum}) and $W$ the total intensity (\ref{eq:WFinal}), and the width of the spectrum as the standard deviation
\begin{equation}
\sigma=\sqrt{\mean{(l-\mean{l})^2}} = \left[\sum_l \frac{(l-\mean{l})^2 W(l)}{W}\right]^{\frac{1}{2}} .
\end{equation}
The results at $qB=-0.2 M^2$, $E=1.5$ and $E= 3 M$ and for $m=-0.5$ and $m=-5.5$ are shown in Fig.~\ref{fig:PAMVsRot}. The data points give the average value $\mean{l}$ and the error bars the deviation $\sigma$. For small rotation we use the unbounded solution (blue line). The red points for $E= 3 M$, $\Omega>0$ and $\rho_R=100$ is the result with unbounded solution but including only final causal state with $n'+a'\leq \rho_R$. As noted earlier, the spectrum has small, negative $\mean{l}$  and narrow width for fast negative rotation, then, as we increase $\Omega$, $\mean{l}$ quickly become positive and for small positive rotation the spectrum rapidly increases in width. For fast positive rotation the average $l$ is lower and distributed much narrower. The green lines show the results for a rotating system in the absence of magnetic field. The angular momentum of photons is directed along the rotation but it does not increase indefinitely. As argued when discussing the intensity of radiation, the reason is that we are considering an initial state with low energy for such a confined state and we do not have enough energy to reach higher angular momentum state, see for instance bottom right plot in Fig.~\ref{fig:energyLevelsJR}.

%*********************************************************************************************************
\section{Summary}\label{sec:summary}
%*********************************************************************************************************
In this paper we performed a detailed analysis of the electromagnetic radiation by a fermion embedded in a uniformly rapidly rotating medium and subject of the external magnetic field. We first obtained the exact solution of the Dirac equation in the finite cylinder with radius $R=c/\Omega$. These states, unlike the ones in an unbounded volume, do not violate causality and the energy spectrum  is bounded from below, ensuring that summations over the complete set of states do not introduce nonphysical divergences. These were essential problems to solve if one wants to study rapidly rotating systems. Between several methods to impose a boundary condition (BC) at the light-cylinder, we used the MIT boundary condition because it guarantees that no conserved currents cross the light-cylinder. Another important difference with the unbounded case is that the integer Landau principal quantum number $n$ becomes a real number whose value depends on other quantum numbers: the total angular momentum along the axis of the cylinder and the polarization, as well as on the size of the cylinder, fermion mass and the magnetic field strength.

Using these causal states we analytically computed the differential radiation intensity given by (\ref{eq:DiffIntensityCylinder}), the spectrum of the photon's angular momentum (\ref{eq:WFinal}) and the total intensity (\ref{eq:Wlspectrum}). The final analytical results still contain summations and integration that we performed numerically. In the unbounded case some of these sums and integrals can be carried out analytically, while in the finite volume that is not possible because the quantization of transverse momentum has to be obtained numerically from the MIT BC constrain (\ref{eq:MITConstraint}) and depends on many parameters.

At slow rotation the synchrotron radiation intensity is suppressed for parallel $q\bm{B}$ and $\bm{\Omega}$ and is enhanced for the antiparallel configuration. Our main observation, illustrated in Fig.~\ref{fig:EnergydepMITBC} and \ref{fig:WVsRot}, is that for fast rotation the effect of enhancement is reduced and that the intensity from the parallel configuration starts to be enhanced as well. The latter occurs simply because the radiation from the global rotation becomes large and dominates over the circular motion caused by the magnetic field. Instead, a likely explanation of the reduced enhancement for the antiparallel configuration is that as the angular velocity increases, the states get squeezed in a smaller volume to preserve causality. This causes all energy levels to rise, and as a result, a state with a given energy has less states to decay into, resulting in a lower emission probability compared to an emission in a larger volume.

We furthermore studied the angular momentum of the radiation. The results illustrated in Fig.~\ref{fig:PAMVsRot}, indicates that the photon angular momentum tends to be aligned in the opposite direction of $q\bm{B}$ and when $\bm{B}$ and $\bm{\Omega}$ are parallel becomes aligned to rotation only at very fast rotation. For moderate rotation the angular momentum is spread almost evenly in a wide range of values including very high values. At faster rotations the angular momentum spread is more peaked toward lower values.

The analysis presented in this work is general and can be applied to any rapidly rotating system. Our main interest and motivation is relativistic heavy-ion collisions, where estimates indicate that the magnetic length and the light-cylinder radius are of similar order of magnitude. The early time magnetic field is on the order of $eB\sim m_\pi^2=2\times 10^4\text{ MeV}^2$. In the quark-gluon plasma phase at temperature $T= 300$ MeV, considering a quark with thermal mass $M\sim T$, we obtain $qB/M^2\sim 0.2$, which is the value used in this work. The late time vorticity of the QGP, which gives the angular velocity of rotation, is measured through spin polarization \cite{STAR:2017ckg}. Contrary to the magnetic field, which is larger at higher collision energies, the vorticity is higher at low energies. Estimates indicate that at low collision energy, the angular velocity $\Omega$ can also be larger than the magnetic field $\sqrt{qB}$, see for instance Fig. 1 of~\cite{Buzzegoli:2023vne}, while at high energies the magnetic field is definitely larger. However, the time evolution suppress the magnetic field much faster than the rotation, which is instead sustained by the conservation of angular momentum, and even in the high collision energy we expect a stage where the magnetic field is lower than the rotation. Thus it emerges that throughout most of plasma evolution $\rho_R=|qB|/{2\Omega^2}\leq 1$, implying that the effects analyzed in this work are expected to be very important.

The numerical computation of the photon radiation by the quark-gluon plasma is quite costly and we leave it for the future study. We would like to point out however, that in the limit of extremely fast rotation, when the light-cylinder radius $1/\Omega$ is much shorter than the mean-free-path, which is in turn much shorter than the plasma size, the calculation significantly simplifies.
First, the magnetic field can  be neglected, as we showed in this paper. Second, we argued in our recent article \cite{Buzzegoli:2023yut} that the electromagnetic spectrum can be computed by considering only the states localized on the light-cylinder surface.  This may be the most realistic path toward the phenomenological modeling of the electromagnetic radiation by rotating quark-gluon plasma.

%%%%%%%%%%%%%%%%%%%%%%%%%%%%%%%%

\acknowledgments
We thank J.D. Kroth and Nandagopal Vijayakumar for many discussions of rotating systems. 
%We thank ... for helpful communications/correspondance.
This work  was supported in part by the U.S. Department of Energy under Grant No.\ DE-SC0023692.

%%%%%%%%%%%%%%%%%%%%%%%%%%%%%%%%%%%%
%
%

%% appendix 
% \appendix

\end{document}